\documentclass[preprint,12pt]{article}
\usepackage[dvips]{graphicx}
\usepackage{ccaption}
\captiondelim {.  } 

\usepackage[centertags]{amsmath}
\usepackage{amsthm,amsfonts,amssymb}
\usepackage{indentfirst}
\usepackage[font={scriptsize}]{caption}

\newcommand{\bfA}{\mathbf{A}}
\newcommand{\bfu}{\mathbf{u}}

\newtheorem{thm}{Theorem}[section]

\title{Fitness Optimization and Evolution of Permanent Replicator Systems}
\date{\vspace{-5ex}}

\author{ 	
	{\bf Sergei Drozhzhin} \\
	 Lomonosov Moscow State University\\ Moscow 119992, Russia\\
	{\bf Tatiana Yakushkina}\\ National Research University Higher School of Economics\\
	 Moscow 101000, Russia\\
	 {\bf Alexander S. Bratus}\\
	 Russian University of Transport\\Moscow 127994, Russia\\
    alexander.bratus@yandex.ru}
	
\begin{document}
	\maketitle
	
\begin{abstract}
In this paper, we discuss the fitness landscape evolution of permanent replicator systems using a hypothesis that the specific time of evolutionary adaptation of the system parameters is much slower than the time of internal evolutionary dynamics. In other words, we suppose that the extreme principle of Darwinian evolution based the Fisher's fundamental theorem of natural selection is valid for the steady-states. Various cases of the evolutionary adaptation for permanent replicator system are considered.  
\end{abstract}

%
%{\bf	model of evolution;\quad replicator system; \quad fitness landscape \quad Fisher's theorem of natural selection \quad permanence }
%	 

\section*{Introduction}
\label{s1}

Starting from Fisher's fundamental theorem of natural selection, evolutionists began to apply extreme principles to Darwinian evolution \cite{par90, gra08, gra06}. The theorem postulates: ``The rate of increase in (mean) fitness of any organism at any time is equal to its genetic variance in fitness at that time'' \cite{fi30}. However, the notion of ``genetic variance of fitness'' was not strictly defined in the early studies.  Later, Wright \cite{wr32}  introduced another important concept~---  adaptive fitness landscape, which is extensively applied in theoretical biology.  

Many of the extreme principles in evolutionary theory rely on the assumptions of the constant fitness landscape and steadily growing mean fitness. In biological studies, underlying understanding of the fitness landscape is often based on common visualization as a statical hypersurface. From this point of view, the evolutionary process is depicted as a path going through the space with hills, canyons, and valleys, ending up in one of the peaks \cite{poel07, birch16}. For the avoidance of doubt and misinterpretation, we define the notion ``\textit{fitness landscape}'' explicitly, providing mathematical formalization for its geometry in the case of general replicator systems.  

Let us start with the classical evolutionary model --- the replicator equation \cite{hz88, sch83, ko12}:
\begin{equation}\label{e1}
\dot{u}_i = u_i \left((\bfA \bfu)_i - f(\bfu)\right), i = 1,\ldots n.
\end{equation}
Here, $\bfA = [a_{ij}]$ is a given $n\times n$ matrix of fitness coefficients.  Value $(\bfA\bfu)_i$ is $i$--th element of the vector $\bfA\bfu,$  where vector $\bfu$ stands for the distribution of the species in the population over time. The term $f$ guarantees that for any time moment $t$ the vector $\bfu(t)$ belongs to the simplex $S_n$:
\begin{eqnarray}\label{e2}
\bfu(t)\in S_n &=& \{\mathbf{x}\in \mathbb{R}^n: x_i\geqslant 0, \sum_{i=1}^{n}x_i =1\},\\
f(\bfu) &=& \sum_{i,j=1}^{n}a_{ij}u_iu_j = \left( \bfA\bfu,\bfu\right).\nonumber
\end{eqnarray}
 In terms of evolutionary game theory, $ f(\bfu)$ is the mean population fitness of the population of composition  $\bfu$ and the payoff matrix $\bfA$. The equilibria ${\bar\bfu}$ are given by the solutions to the system of algebraic equations:
\begin{equation}\label{e3}
\bfA\bar{\bfu} = {f}(\bar{\bfu})\mathbf{1},\quad \mathbf{1} = (1,\ldots, 1)^T, \bar{\bfu}\in S_n,
\end{equation}
where ${f}(\bar{\bfu})$ is the mean fitness at the equilibrium state $\bar{\bfu}$ (which is not necessarily stable).

\textit{
	The geometric object $\Sigma$ in $\mathbb{R}^n$  corresponding to the following quadratic form:
	\begin{equation}\label{e4}
		f(\bfu) = \sum_{i,j =1}^{n} a_{ij}u_iu_j =  \left( \bfA\bfu,\bfu\right),
	\end{equation}
	where $\bfu(t)\in S_n = \{\bfu \in \mathbb{R}^n: \forall i\quad u_i\geqslant 0, \sum_{i=1}^{n}u_i =1\},$
	defines the \textbf{fitness landscape} of the system (\ref{e1}). 
}

Note that for every trajectory $\gamma_t$ of the system (\ref{e1}), $\gamma_t\in S_n$, there is a  curve $\Gamma_t\in \Sigma$. Since any matrix $\bfA$ can be decomposed into the sum:
$$
\bfA = \frac{1}{2}(\bfA+\bfA^T) + \frac{1}{2}(\bfA-\bfA^T) = \mathbf{B}+\mathbf{C},
$$ 
where $\mathbf{B}$ is a symmetric matrix, $\mathbf{C}$  is a skew-symmetric matrix, then 
$$
f(\bfu) = \left(\mathbf{B}\bfu, \bfu\right).
$$
This means that there is an orthogonal transformation $\mathbf{U}$, such that 
$$
\mathbf{U}^T\mathbf{B}\mathbf{U} = \Lambda = diag(\lambda_1, \ldots, \lambda_n).
$$
Here, the values $\lambda_i$ stand for real eigenvalues of the matrix $\mathbf{B}$. Hence, the transformation $\bfu = \mathbf{U}\mathbf{w}$ reducts the quadratic form to the canonical form:
\begin{equation}\label{e5}
	f(\mathbf{w}) = \sum_{i=1}^{k}\lambda^+_iw_i^2 + \sum_{j=k+1}^{n}\lambda_j^-w^2_j,
\end{equation} 
where $\lambda_i^+$ and $\lambda_j^-$ denote positive and negative real eigenvalues of $\mathbf{B}$ correspondingly (assuming that $|\mathbf{B}|\neq0$). The same transformation affects the simplex $S_n$, giving the convex set:
\begin{equation}\label{e6}
	W_n = \{\mathbf{w}\in \mathbb{R}^n, \left(\mathbf{w}, \mathbf{U}^T\mathbf{1}\right) =1, \mathbf{Uw}\geqslant 0, \mathbf{1} = (1,\ldots, 1)^T\in \mathbb{R}^n\}.
\end{equation}
 Thus, the fitness landscape of the general replicator system (\ref{e1}) is defined by the shape of the surface (\ref{e5}), which has a canonical form for a convex set (\ref{e6}). 

We suggest extending classification for quadratic forms on fitness landscapes, emphasizing their geometrical features. Thus, we define three types of fitness landscapes, depending on the eigenvalues of the matrix $\mathbf{B}$:  \textit{elliptic} if all the eigenvalues are of the same sign, \textit{hyperbolic} if some eigenvalues have opposite signs, and \textit{parabolic} if there are zero eigenvalues. Autocatalytic replicator systems have elliptic fitness landscapes with $n$ peaks in the corners of the simplex $S_n$, each of which is an attractor. The trajectories $\gamma_t$, depending on the initial state, belong to different basins of attraction, and the curves $\Gamma_t$ converge to one of the corresponding attractors.  In hypercycle systems with the coefficients $k_i =1, i=1,\ldots,n$, the eigenvalues of the matrix $\mathbf{B}$ can be obtained as $\lambda_k = \cos\frac{2\pi}{n}k, k=0,\ldots n-1$ (where $\lambda_0$ corresponds to the vector $(1, \ldots, n)$, which does not belong to the simplex $S_n$). In this case, the fitness landscape has a hyperbolic type. The hyperspace $\Sigma$ reaches its maximum heights on the simplex border for $n\geqslant 5$ when the system has a limit cycle, and in a steady-state for $n=2,3,4,$ when there is a stable equilibrium. These illustrate the correspondence between $\Sigma$ geometry and the behavior of phase trajectories. 

From a mathematical perspective, Fisher's fundamental theorem is correct only for systems with symmetric matrices of interaction, which corresponds to the diploid population. Moreover, from these assumptions, the maximum fitness value should be reached in the steady-state of the evolutionary system. This forms significant restrictions on the applicability, making these cases rather exceptional than realistic \cite{bra18}. Various studies \cite{ew89, le97,ao05, ao09, ew15} were dedicated to new interpretations and re-consideration of  Fisher's postulates. In \cite{ao05}, e.g., which provides an extensive literature review on mathematical formalism for the fundamental laws in evolution,   Fisher's approach to natural selection is discussed in terms of the F-theorem. In the current study, we develop a fitness optimization technique introduced in \cite{drozh18}.

\subsection*{Dynamical fitness landscapes: adaptation process}
One of the ways to examine fitness landscapes is to consider their fluctuations. The question arises: how the adaptive changes can be achieved in evolution. The central hypothesis of this study is that the specific time of the evolutionary adaptation of the system parameters is much slower than the time of the internal evolutionary process, which leads the system to its steady-state. Throughout the paper, we will call the first the evolutionary time. For hypercycles, we introduced a similar concept in the study  \cite{drozh18}. This assumption leads to the fact that evolutionary changes of the system parameters happen in a steady-state of the corresponding dynamical system.

In other words, we can write an equation for a steady-state with respect to the evolutionary time over a set of possible fitness landscapes. Consider a population distribution $\bfu = (u_1, \ldots, u_n)$ representing the frequencies of different types in a replicator system. If the system is permanent over a simplex $S_n$ (here, the notation is the same as in the previous subsection) and there is a unique internal equilibrium $\mathbf{u}\in int S_n$ (stable or unstable one), then the mean integral value of the frequencies and the mean fitness value coincide with ones that reached in a steady-state. This allows examining an evolutionary process of fitness landscape adaptation using only the equation for a steady-state, where all the elements depend on the evolutionary time $\tau$. Therefore, fitness landscape adaptation happens in time, which describes system dynamics converging to a steady-state over the set of possible fitness landscapes. 

It is worth pointing out, that this approach is valid only for permanent systems. In this case, it holds:
\begin{eqnarray}\label{e7}
	\bar{u}_i = \lim_{T\rightarrow +\infty}\frac{1}{T}\int_{0}^{T}u_i(t)dt, \nonumber\\
	\bar{f} = \lim_{T\rightarrow +\infty}\frac{1}{T}\int_{0}^{T}\left( \bfA\bfu(t),\bfu(t)\right) dt.
\end{eqnarray}
The adoptation of the fitness landscape over the evolutionary time $\tau$ can be described by:
\begin{equation}\label{e8}
\bfA(\tau)\bar{\bfu}(\tau) = \bar{f}(\bar{\bfu}(\tau))\mathbf{1}, \quad \bar{\bfu}(\tau)\in S_n,
\end{equation}
where $a_{ij}((\tau))$ of the matrix $\bfA$ are smooth functions with respect to the parameter $\tau\in [0,+\infty)$. Each solution of the equation (\ref{e8}) corresponds to the dynamics of the permanent replicator system, which is characterized by (\ref{e7}).

The possible fitness landscapes satisfy the condition:
\begin{equation}\label{e9}
	\|\bfA(\tau)\|^2 = \sum_{i,j=1}^{n}a_{ij}^2 \leqslant Q^2, \quad\tau\geqslant 0, Q = const. 
\end{equation}
We show that the problem of evolutionary adoptation of the replicator system over the set of possible fitness landscapes (\ref{e9}) transforms into the problem of function $\bar{f}(\tau)$ maximization over the set of the solutions to (\ref{e8}).

In a previous study \cite{drozh18}, we obtained an expression for the mean fitness variation. Based on this result, we suggested a process of fitness optimization in the form of a linear programming problem. The numerical simulations show that during the iteration process, the fitness value increases and the systems behaviour changes. For the big enough number of iterations, the steady-state of the system stays almost the same; however, the fitness value grows drastically at the same time. In this case, when the initial state of the system is described by the hypercycle equations, we see a qualitative transformation of the system: new connections appear, and autocatalysis can start. Increasing the number of iterations further, the coordinates of the steady-states split and, over the simplex, the system converges to a fixed fitness value. In \cite{drozh18}, it was shown, that the observed effect is similar to an ``error threshold'' effect in the quasispecies system by Eigen \cite{ei71, ei79}, when the eigenvalue and the eigenvector of the system stay unchanged with the mutation rate growth. 

In the Appendix, we provide mathematical proof for this effect. That is, if the mean fitness  reaches the extrema for some evolutionary parameter's value, then the mean fitness does not grow further even for larger parameter values.  

The main question is the relevance of the approach to the ESS (evolutionarily stable strategy)
concept, which is widely used in the studies focused on the evolutionary game theory. 
It is known, that if the state $\bfu(\tau)\in intS_n$ is evolutionary stable, then it is asymptotically stable \cite{ho98}. One can also show, that in this case the steady-state is a local extrema point for the mean fitness function \cite{bra18}.

Consider a special case of the replicator system --- a hypercycle system: 
\begin{eqnarray}\label{e10}
&&\dot u_i = u_i\left(k_iu_{i-1} - f(t)\right),\quad i=1,\ldots,n, \bfu\in S_n,\nonumber\\
&&f(t) = \sum_{i=1}^{n}k_iu_iu_{i-1}, \quad u_0 = u_n.
\end{eqnarray}
For $n\geqslant 5$, there is no evolutionary stable state, since the only equilibria $\hat \bfu \in intS_n$
is unstable and a stable limit cycle exists.  Despite this fact, the suggested process of evolutionary adaptation of the fitness landscape can drastically change the mean fitness of the system (\ref{e10}) without affecting the coordinates of the steady-state. In the case of hypercycles with small size ($n=2,3$), the steady-state is ESS. The numerical simulations show, that every step of the iteration process of the evolutionary adaptation this ESS property holds and the fitness function steadily growth (see Fig.1).

%\begin{figure}[h!]
%	\center
%	\includegraphics[scale=0.25]{1.eps}
%%	\includegraphics[scale=0.2]{2.eps}
%	\caption{The evolutionary dynamics of the mean fitness of the hypercycle system (6) n=3 changing over evolutionary time $\tau$}
%\end{figure}

In this paper, the approach of the evolutionary adaptation process is applied to different classes of replicator systems. In the first part, we consider the system of cyclic replication, where each species depends on the two previous ones in the scheme. We call this class  ``bi-hypercycles'' or binary hypercycles. In the second part, the evolutionary process targets the case, when at  any random time moment a new species can be added to a hypercycle. The main goal here is to examine how the combination of determined and stochastic factors affects the replicator system's evolutionary  
dynamics. The importance of this direction was highly recognized in the literature \cite{ko12}.
In the last part of the paper, we consider two specific cases. First, we analyze a hypercycle with a dominating type, which influences all the evolutionary process. We call this class of the replicator systems ``ant hill'' in association with the internal dynamics of ant populations. In another example, we study a specific matrix of transition for the replicator system inspired by the experimental results \cite{vm12}.

\section*{Evolution of Bi-hypercycles}
Let us consider the following system, which we call a ``bi-hypercycle'' system throughout the paper:
\begin{eqnarray}\label{e11}
	&&\dot u_{i}(t) = u_{i}(t)\Big(a_{i}a_{i - 1}u_{i - 1}(t)u_{i - 2}(t) - f(t)\Big), \quad i = 1, \ldots, n,\nonumber\\
	&&f(t) = \sum\limits_{i = 1}^{n}a_{i}a_{i - 1}u_{i}(t)u_{i - 1}(t)u_{i - 2}(t).
\end{eqnarray}
The reproduction of each element in the hypercycle of this type is catalyzed by two previous elements. 
As in the previous models, the frequencies of the types belong to a simplex:
\begin{equation}\label{e12}
S_{n} = \Big\{\bfu: u_{i} \geqslant 0, \sum\limits_{i = 1}^{n}u_{i} = 1\Big\},
\end{equation}
where $a_{i} > 0,$ $a_{0} = a_{n},$ $u_{0}(t) = u_{n}(t), u_{-1}(t) = u_{n - 1}(t)$.

We rewrite the equations (\ref{e11}) for ${\bf u}(t) = (u_{1}(t), u_{2}(t),\ldots, u_{n}(t))$ in a matrix form. For doing this, we introduce 
the transition matrix:
\begin{equation}\label{e13}
{\bf A} = \left(
\begin{array}{ccccc}
0 & 0 & \ldots & 0 & a_{1}\\
a_{2} & 0 & \ldots & 0 & 0\\
0 & a_{3} & \ldots & 0 & 0\\
. & . & . & . & .\\
0 & 0 & \ldots & a_{n} & 0\\
\end{array} \right).
\end{equation}
To finalise the matrix equation, we denote:
$$
{\bf U(t)} = diag(u_{1}(t), u_{2}(t), \ldots, u_{n}(t)), \quad {\bf 1} = (1, 1, \ldots, 1)^T.
$$
The equalities (\ref{e11}, \ref{e12}) transform into:
\begin{eqnarray}\label{e14}
&&\dot u_{i}(t) = u_{i}(t)\Big(\left(\left({\bf AU}(t)\right)^{2}{\bf 1}\right)_{i} - f(t)\Big), \quad i = 1, \ldots, n,\nonumber\\
&&f(t) = \Big(\left({\bf AU}(t)\right)^{2}{\bf 1}, {\bf u}(t)\Big), \nonumber\\
&&\Big({\bf U}(t){\bf 1}, {\bf 1}\Big) = 1. 
\end{eqnarray}
The system (\ref{e11}, \ref{e12}) is permanent, hence \cite{safro13} the values of the matrix ${\bf U}(t)$ and function $f$ are defined in a steady-state as follows:
\begin{eqnarray}\label{e15}
\bar{\bf {U}} = \lim_{T \to \infty}\frac{1}{T}\int\limits_{0}^{t}{\bf U}(t)dt,\nonumber\\
\bar{f} = \lim_{T \to \infty}\frac{1}{T}\int\limits_{0}^{t} f(t)dt.
\end{eqnarray}
Consider the set of non-negative matrices ${\bf A}(\tau) = (a_{kj}(\tau))_{k,j = 1}^{n}$, where elements $a_{kj}$ are smooth functions with respect to $\tau$. The condition (\ref{e9}) applies here:
\begin{equation}\label{e16}
\sum\limits_{k,j = 1}^{n} a_{kj}^{2}(\tau) \leqslant Q, \quad Q = const > 0. 
\end{equation}
Moreover, we assume ${\bf A}(0) = {\bf A}$, where ${\bf A}$ --- is a matrix (\ref{e13}).

We apply the  hypothesis concerning evolutionary changes of the matrix elements ${\bf A}(\tau)$ 
during the fitness optimization process. The steady-state is described by the equality:
\begin{equation}\label{e17}
\Big({\bf A\bar{U}}\Big)^{2}{\bf 1} = \bar{f}{\bf 1}, \quad \Big({\bf \bar{U}}{\bf 1}, {\bf 1}\Big) = 1. 
\end{equation}
From here it follows that the equation for evolutionary adaptation of the system has the form:
\begin{equation}\label{e18}
\Big({\bf A}(\tau){\bf \bar{U}}(\tau)\Big)^{2}{\bf 1} = \bar{f}(\tau){\bf 1}, \quad \Big({\bf \bar{U}}(\tau){\bf 1}, {\bf 1}\Big) = 1. 
\end{equation}
Let us vary the evolutionary time parameter $\tau$ to $\tau + \Delta \tau$ and examine the perturbation of the matrix ${\bf A}(\tau), {\bf \bar{U}}(\tau)$, vector  ${\bf \bar{u}}(\tau)$, and function $\bar{f}(\tau)$. We denote as $\delta {\bf A}, \delta{\bf \bar{U}}, \delta\bar{u}$, and $\delta\bar{f}$ the corresponding linear parts of the increments. From (\ref{e18}), we get the equation with the accuracy $o(\Delta\tau)$ :
\begin{eqnarray}\label{e19}
\Big(\left({\bf A} + \delta{\bf A}\right)\left({\bf \bar{U}} + \delta{\bf \bar{U}}\right)\Big)^{2}{\bf 1} = \bar{f}{\bf 1} + {\bf A\bar{U}A}\Big(\delta{\bf \bar{u}}\Big) +\nonumber \\  + {\bf A\bar{U}}\Big(\delta{\bf A}\Big){\bf \bar{u}} + {\bf A}\Big(\delta{\bf \bar{U}}\Big){\bf A\bar{u}} + \Big(\delta{\bf A}\Big){\bf \bar{U}A\bar{u}}.
\end{eqnarray}
Let  ${\bf \bar{v}} = (v_{1}, v_{2}, \ldots, v_{n}) = {\bf A\bar{u}},$ ${\bf V} = diag(v_{1}, v_{2}, \ldots, v_{n})$. Since
$$
\Big(\delta{\bf \bar{U}}\Big){\bf A\bar{u}} = \Big(\delta{\bf U}\Big){\bf \bar{v}} = {\bf V}\Big(\delta{\bf \bar{u}}\Big),
$$
then
\begin{equation}\label{e20}
\Big(\delta\bar{f}\Big){\bf 1} = {\bf A}\Big({\bf \bar{U}A} + {\bf V}\Big)\delta{\bf \bar{u}} + \Bigg({\bf A\bar{U}}\Big(\delta{\bf A}\Big) + \Big(\delta{\bf A}\Big){\bf \bar{U}A}\Bigg){\bf \bar{u}}. 
\end{equation}
Multiplying the latter equality by the vector
$$
\Bigg(\Big({\bf A\bar{U}A} + {\bf AV}\Big)^{-1}\Bigg)^{T}{\bf 1},
$$
and taking into account
$$
\Bigg(\Big(\delta{\bf \bar{U}}\Big){\bf 1}, {\bf 1}\Bigg) = 0,
$$
we get the formula for the mean fitness variation with respect to the small parameter perturbation $\tau$:
\begin{equation}\label{e21}
\delta\bar{f} = \frac{\Bigg(\Big({\bf \bar{U}A} + {\bf V}\Big)^{-1}\Big({\bf A}^{-1}(\delta{\bf A}){\bf \bar{U}A} + {\bf \bar{U}}(\delta{\bf A})\Big){\bf \bar{u}}, {\bf 1}\Bigg)}{\Bigg(\Big({\bf \bar{U}A} + {\bf V}\Big)^{-1}{\bf A}^{-1}{\bf 1}, {\bf 1}\Bigg)}. 
\end{equation}

According to (\ref{e16}), elements of the matrix  $\delta{\bf A}(\tau)$ have to satisfy the condition 
\begin{equation}\label{e22}
\sum\limits_{i, j = 1}^{n}a_{ij}(\tau)\delta a_{ij}(\tau) \leqslant 0.
\end{equation}
The correspondence of the value   $\delta{\bf \bar{u}}(\tau)$ to the matrix elements  $\delta {\bf A}(\tau)$ can be obtained from (\ref{e20}):
\begin{equation}\label{e23}
\delta {\bf \bar{u}} = \Big({\bf \bar{U}A} + {\bf V}\Big)^{-1}{\bf A}^{-1}\Bigg(\Big(\delta \bar{f}\Big){\bf 1} - \Big({\bf A\bar{U}}\Big(\delta{\bf A}\Big) + \Big(\delta{\bf A}\Big){\bf \bar{U}A}\Big){\bf \bar{u}}\Bigg).
\end{equation}
Note that the condition $\delta {\bf \bar{u}} > 0$ has to be taken into account, if one or several components  ${\bf \bar{u}}$ are close to the simplex' border $S_{n}$.

The set of matrices, falling under the condition (\ref{e16}) is convex. Moreover, the set  ${\bf \bar{u}} \in int S_{n}$ is convex. However, this does not guarantee that the  function $\bar{f}(\tau)$ in (\ref{e18}) is convex. 

Let us find the necessary conditions for the extremum. From (\ref{e4}), we get:
\begin{equation}\label{a1}
(\delta \bfA){\bf \bar{u}} + \bfA(\delta {\bf \bar{u}}) = (\delta \bar{f}){\bf 1}.
\end{equation}
Here, the matrix $(\delta \bfA)$ has elements $a'_{ij}(\tau), i,j = 1,\ldots n,$ and the vector $\delta \bf \bar{u}$ has components:
$$
\delta u_i (\tau) = u'_i(\tau), i =1,\ldots,n,
$$
as well as
$$
\delta \bar{f} = f'(\tau).
$$
Consider the following equation:
$$
\bfA^T(\tau){\bf \bar{v}} = \bf 1, {\bf \bar{v}} = (v_1, \ldots, v_n).
$$
Since $\bfu \in S_n,$ then $({\bf \bar{u}}, \bf 1) = 1$, hence $(\delta {\bf \bar{u}}, \bf 1) = 0.$

Multiplying (\ref{a1}) by ${\bf \bar{v}}$, we obtain:
$$
\left((\delta \bfA){\bf \bar{u}}, {\bf \bar{v}}\right) + \left(\delta {\bf \bar{u}}, \bf 1\right) = \left(I, {\bf \bar{v}}\right)\delta \bar{f}. 
$$
From here, we have the expressions:
$$
\delta \bar{f} = \frac{\left((\delta \bfA){\bf \bar{u}}, {\bf \bar{v}}\right)}{\left({\bf \bar{v}}, \bf 1\right)},
$$ 
$$
\left({\bf \bar{v}}, \bf 1\right) = \frac{1}{\bar f}>0.
$$
If the inequality (\ref{e22}) reaches an equality, then the necessary condition for extremum takes the form:
\begin{equation}\label{a2}
u_i(\tau)v_i(\tau) = \mu a_{ij}, i, j = 1, \ldots, n, \mu = const.
\end{equation}
For the cases where (\ref{a2}) takes place, we observe the effect similar to the ``error threshold'' phenomena in the quasispecies models \cite{ei71}. In \ref{app}, we derive the maximum condition for the mean fitness value , which can be described as an inequality $\delta^2 f(\tau)<0$
for $\delta f(\tau) =0.$

Expression (\ref{e21}) shows the iteration process of the mean fitness maximization, where each step is a linear programming problem. 
As the matrix elements ${\bf A}(\tau)$ are smooth, then
\begin{equation}\label{e24}
a_{ij}(\tau + \Delta \tau) = a_{ij}(\tau) + a_{ij}^{'}(\tau)\Delta\tau + \frac{1}{2}a_{ij}^{''}(\tau)\Delta\tau^{2} + o(\Delta \tau^{2}). 
\end{equation} 
Based on the hypothesis that the elements of the fitness landscape change slowly, suppose that for  $\varepsilon > 0$. We have:
$$
|a_{ij}^{'}(\tau)| \leqslant \varepsilon, \quad |a_{ij}^{''}(\tau)| \leqslant \varepsilon.
$$
If we put $\Delta \tau = \varepsilon$ in (\ref{e24}), then the error for each step of the iteration process, which is using only linear approximations, will be of order $o(\varepsilon^{2})$.
As an initial state ($\tau = 0$), we consider the steady-state (\ref{e17}). Using (\ref{e16}), we solve the linear programming problem:
\begin{equation}\label{e25}
\left\{
\begin{array}{rcl}
\delta\bar{f}(0) = \bar{f}^{'}(0) &\to& \max,\\
\sum\limits_{i, j = 1}^{n}a_{ij}(0)a_{ij}^{'}(0) &\leqslant& 0.
\end{array}
\right.
\end{equation}
As a result, we have such perturbations of the matrix ${\bf A}$, that guarantee the mean fitness growth with the evolutionary time change $\Delta\tau$. The matrix $\delta {\bf A}(0)$ is used for the mean fitness calculation on the next step. Thus, each step includes the choice of the optimal matrix over the matrix set  (\ref{e16}), which leads to the increasing fitness value.

\subsection*{Numerical calculations for bi-hypercycle: the result of the iteration algorithm of the evolutionary process with $n=5$}
As an example of the bi-hypercycle system (\ref{e11}, \ref{e12}), we take $n=5$ with the transition matrix:
\begin{equation}\label{e26}
{\bf A} = \left(
\begin{array}{ccccc}
0 & 0 & 0 & 0 & 1\\
1 & 0 & 0 & 0 & 0\\
0 & 1 & 0 & 0 & 0\\
0 & 0 & 1 & 0 & 0\\
0 & 0 & 0 & 1 & 0\\
\end{array}\right),
\end{equation}
Taking the matrix (\ref{e26}) as the initial condition $\tau =0$, we apply the adaptation process described above and calculate the dynamics of the system (\ref{e11}) numerically. 

The main properties of the system (\ref{e11}, \ref{e26}) are investigated numerically and illustrated by the figures below. Figure \ref{fig1} depicts the evolutionary change of the mean fitness in the timescale $\tau$. The graph for $\bar{f}$ represents the convergence of the fitness to the maximum value after some time moment. As figure \ref{fig2} shows, the position of the steady-state remains the same over around 300 steps of the evolutionary process. After some critical time, the equilibrium point splits into different trajectories: one of the coordinates converges to $1$, while the other four elements --- to $\delta$, which is set up in the algorithm. Figure \ref{fig3} represents the bi-hypercycle system evolving after 200 steps of the adaptation process in the regular timescale $t$, where the stable equilibrium is shown $\bar u_i =0.2,$ $i=1,2,\ldots, 5.$ At the same timescale, Figure \ref{fig4}
shows the dynamics of the bi-hypercycle system, but for 450-th step: unlike the case shown in Figure \ref{fig4}, here we do not see any stable equilibria, and the trajectories have cycles.  The transitions within the system are represented in Fig. \ref{fig5}.
\begin{figure}[h!]
	\centering
	\includegraphics[scale=0.17]{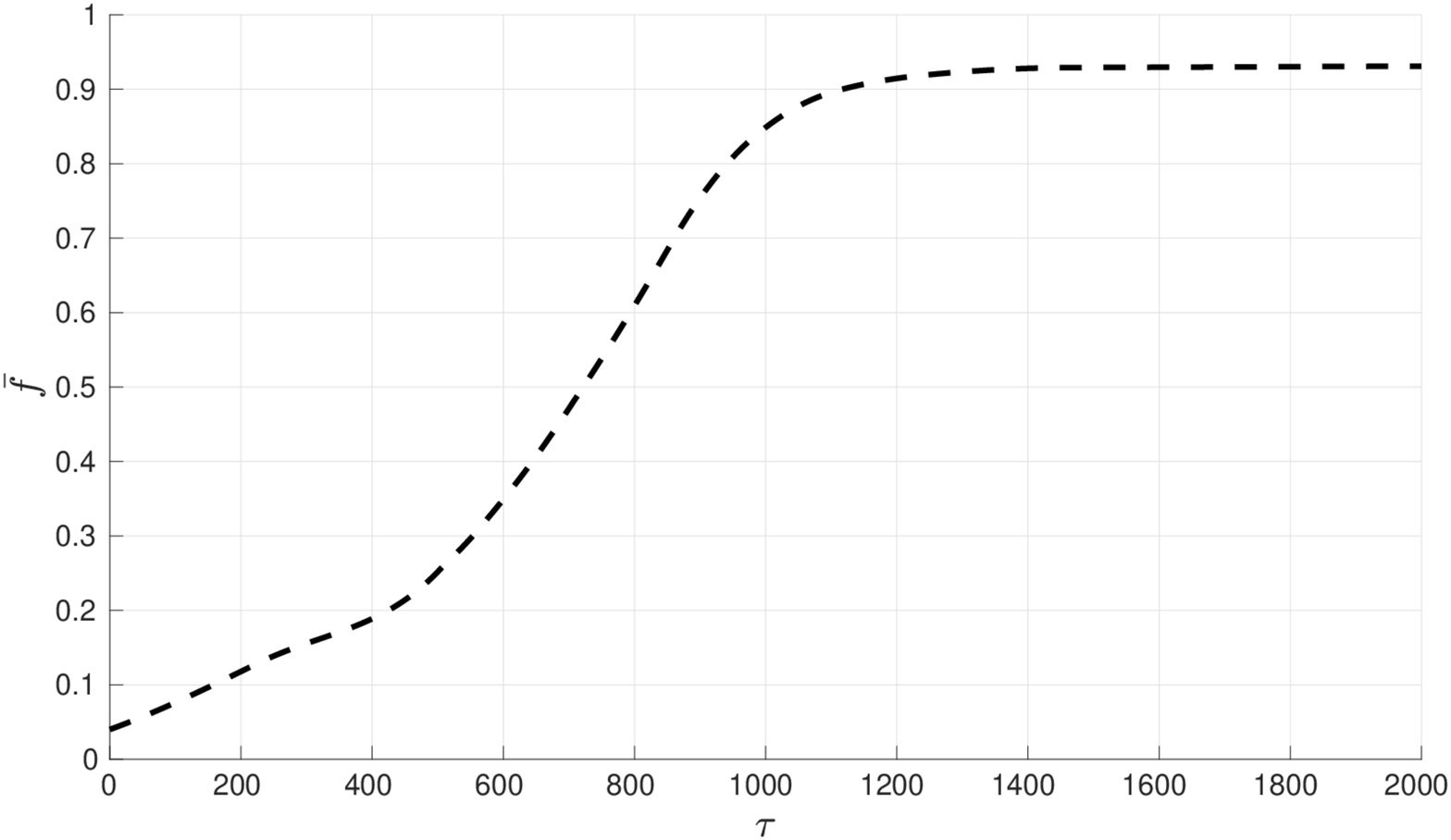}
	\caption{The mean fitness $\bar{f}$ of the bi-hypercycle system (\ref{e11}) with matrix {\bf A} (\ref{e26}) changing over the evolutionary time $\tau$, $n = 5$}\label{fig1}
\end{figure}

\begin{figure}[h!]
	\centering
	\includegraphics[scale=0.17]{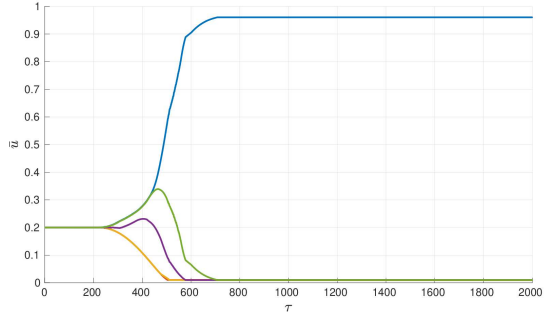}
	\caption{Steady-state ${\bf u} = \bar{u}$ of the bi-hypercycle system (\ref{e11}) with matrix {\bf A} (\ref{e26}) changing over the evolutionary time $\tau$, $n = 5$}\label{fig2}
\end{figure}

\begin{figure}[h!]
	\centering
	\includegraphics[scale=0.17]{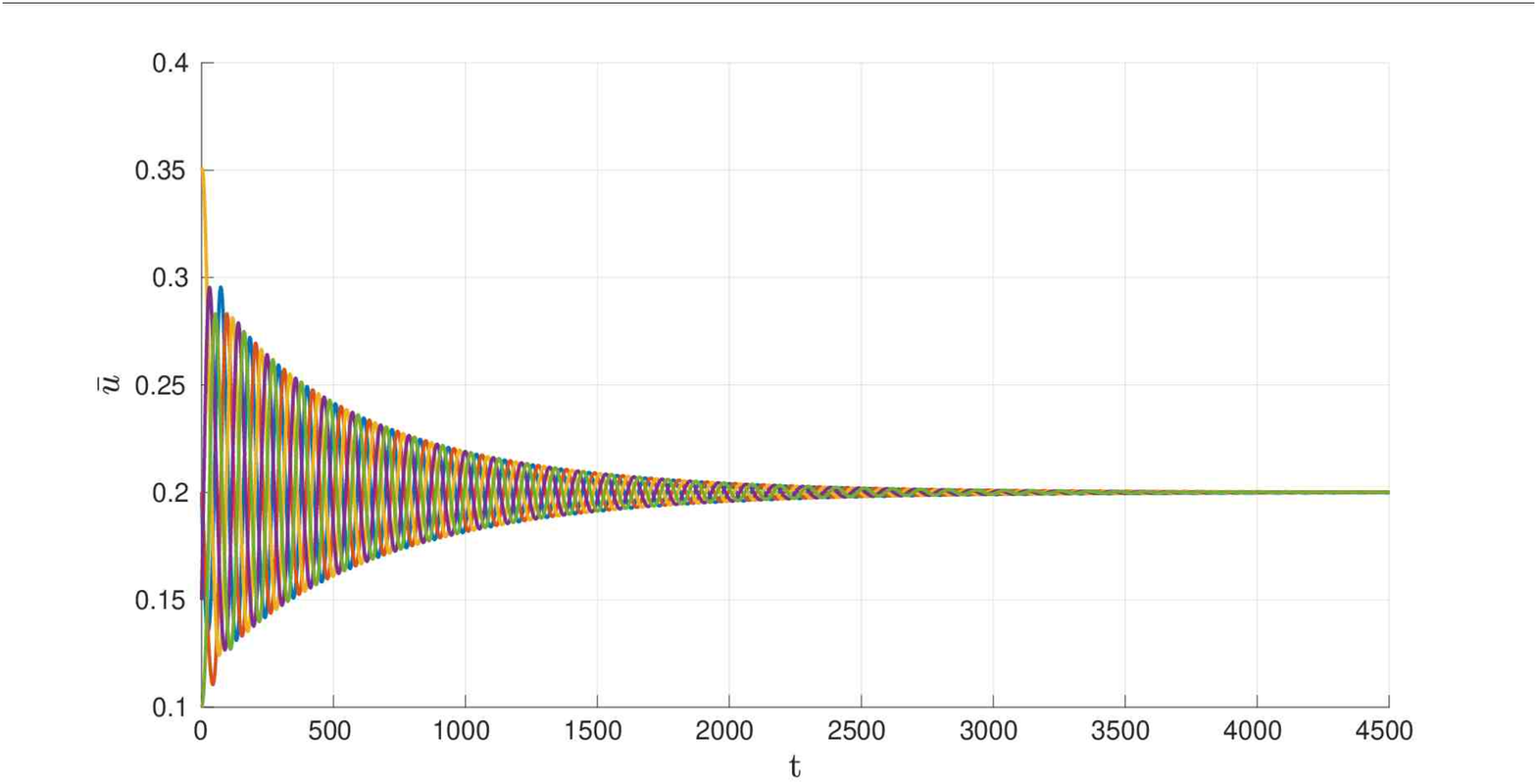}
	\caption{Frequencies of the species in the bi-hypercycle system (\ref{e11}) with matrix {\bf A} (\ref{e26}) changing over the system time $t$ at the 200-th step of the fitness landscape evolutionary process, $n = 5$}\label{fig3}
\end{figure}

\begin{figure}[h!]
	\centering
	\includegraphics[scale=0.17]{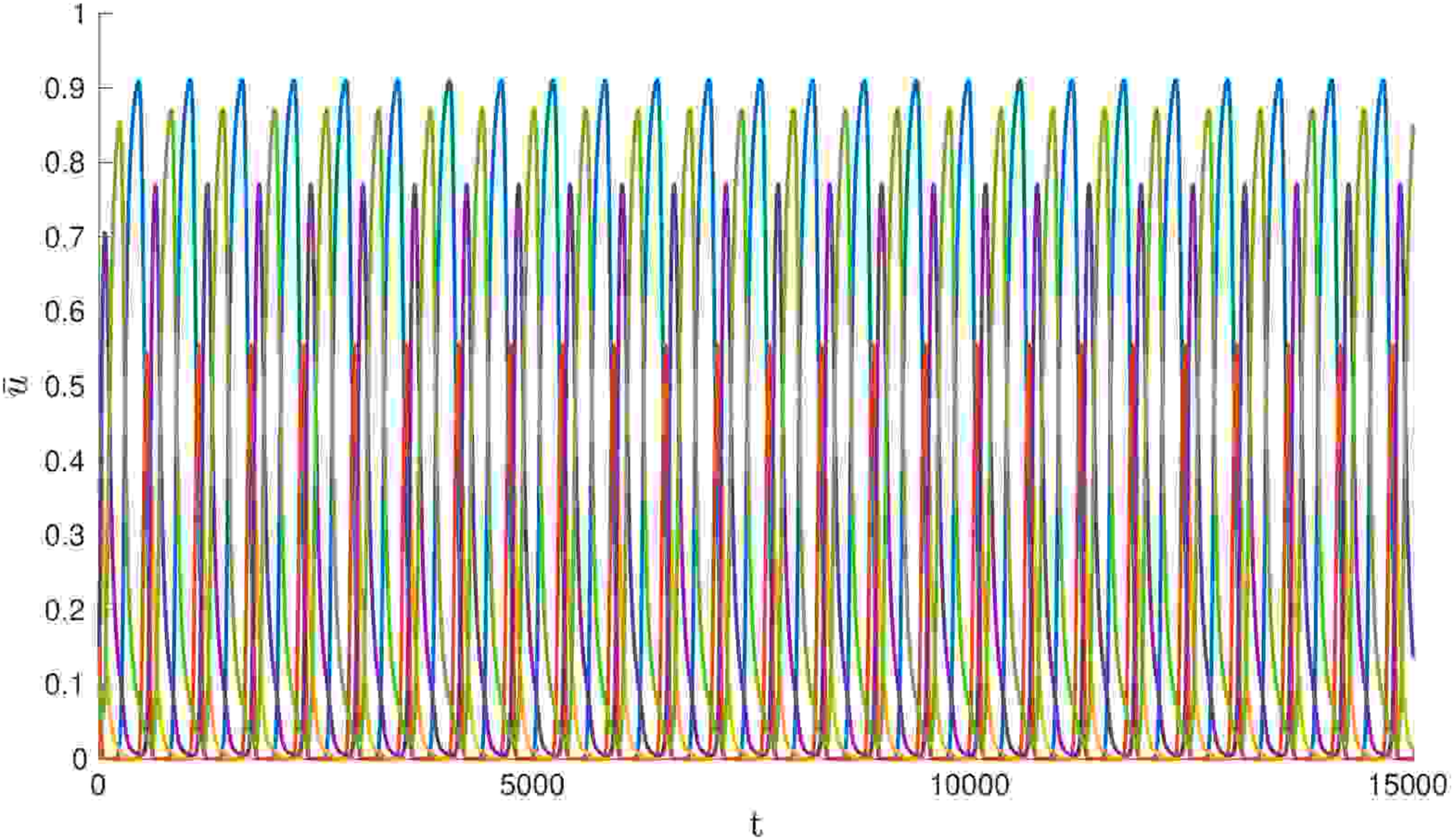}
	\caption{Frequencies of the species in the bi-hypercycle system (\ref{e11}) with matrix {\bf A} (\ref{e26}) changing over the system time $t$ at the 450-th step of the fitness landscape evolutionary process, $n = 5$}\label{fig4}
\end{figure}

\begin{figure}[h!]
	\begin{minipage}[h]{0.47\linewidth}
		\center{\includegraphics[width=0.87\linewidth]{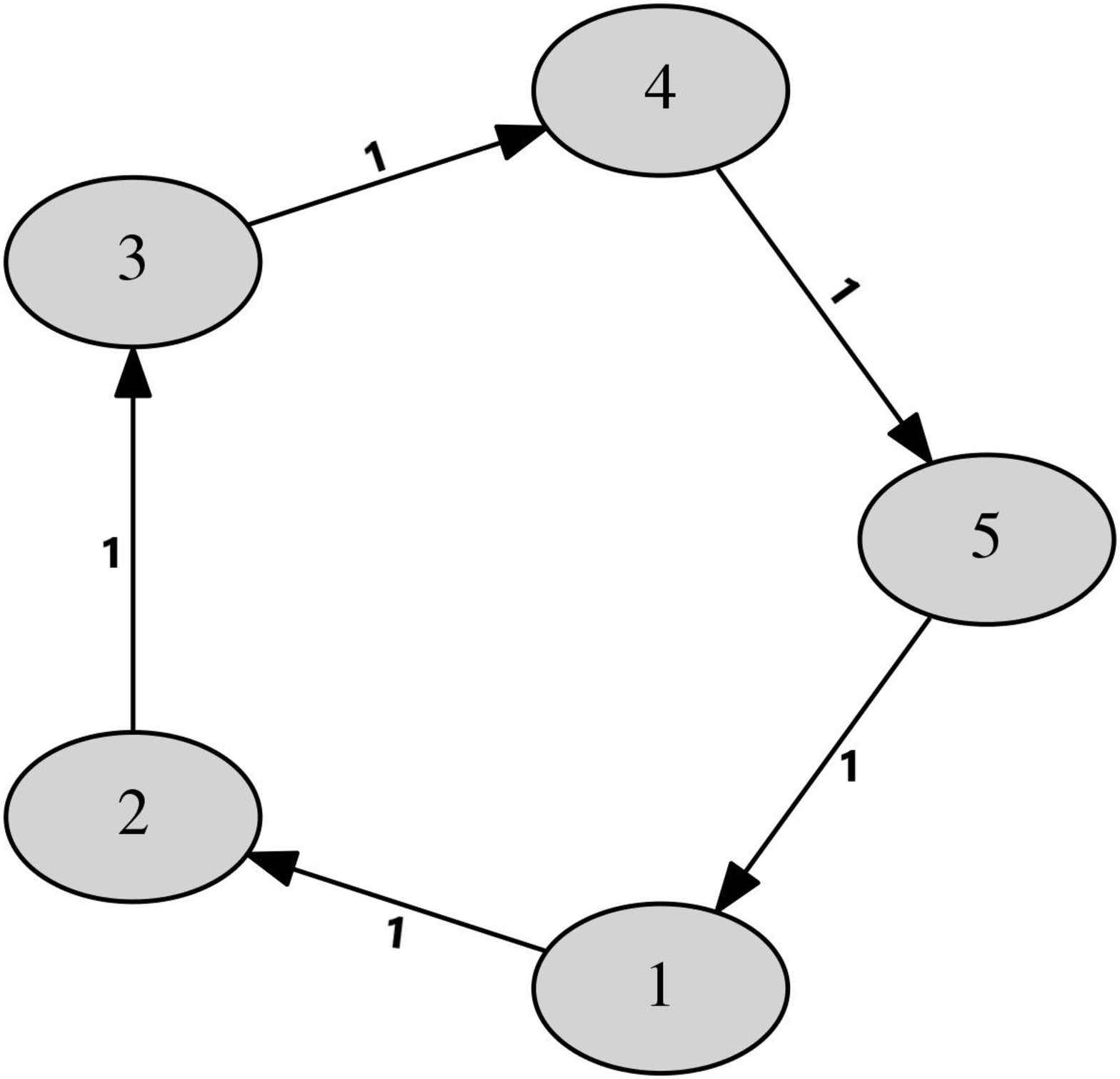} \\ a)}
	\end{minipage}
	\hfill
	\begin{minipage}[h]{0.47\linewidth}
		\center{\includegraphics[width=0.87\linewidth]{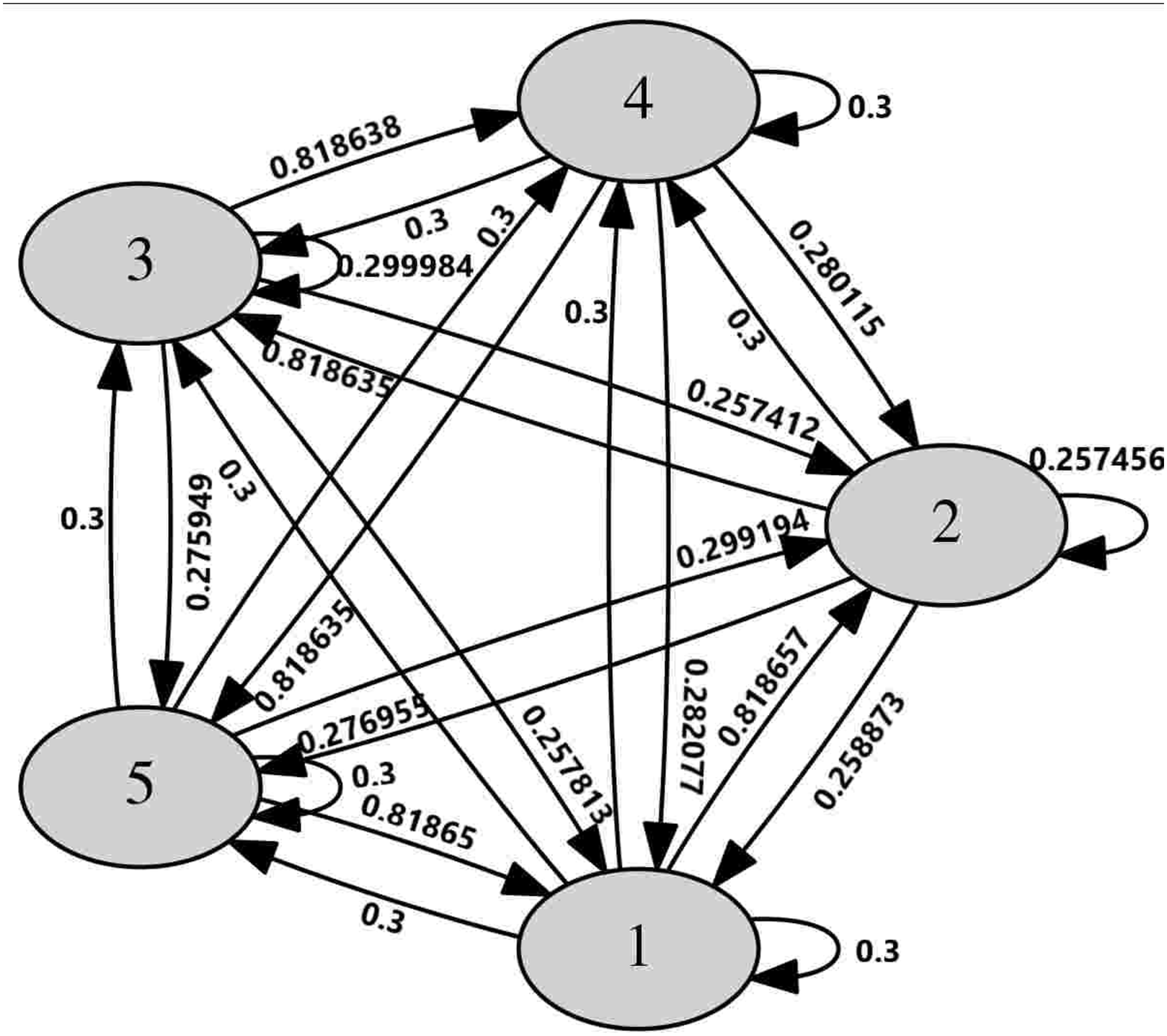} \\ b)}
	\end{minipage}
	\caption{a) Graph represents the original bi-hypercycles before the evolutionary process. Numbers on the edges correspond to the fitness coefficients. b) Graph represents the evolved bi-hypercycles: the result of the evolutionary process after 300 iterations. Numbers on the edges correspond to the adapted fitness coefficients}\label{fig5}
\end{figure}
\newpage
\subsection*{Numerical calculations for bi-hypercycle: the result of the iteration algorithm of the evolutionary process with two parasites}
 Consider another example of the bi-hypercycle system (\ref{e11}, \ref{e12}), we take the transition matrix of the form:
\begin{equation}\label{e27}
{\bf A} = 
\left(\begin{array}{ccccccc}
0 & 0 & 0 & 0 & 1 & 0 & 0\\
1 & 0 & 0 & 0 & 0 & 0 & 0\\
0 & 1 & 0 & 0 & 0 & 0 & 0\\
0 & 0 & 1 & 0 & 0 & 0 & 0\\
0 & 0 & 0 & 1 & 0 & 0 & 0\\
0 & 0 & 0 & 1.1 & 0 & 0 & 0\\
0 & 0 & 0 & 0 & 1.1 & 0 & 0\\
\end{array}\right).
\end{equation}
After the iteration process of evolutionary adaptation, the matrix transforms:
\begin{equation}\label{e28}
{\bf A} = 
\left(\begin{array}{ccccccc}
0.03 & 0.03 & 0.03 & 0.03 & 0.998096 & 0 & 0\\
0.999063 & 0.03 & 0.03 & 0.03 & 0.03 & 0 & 0\\
0.03 & 0.998857 & 0.03 & 0.03 & 0.03 & 0 & 0\\
0.03 & 0.03 & 0.997774 & 0.03 & 0.03 & 0 & 0\\
0.03 & 0.03 & 0.03 & 0.997564 & 0.03 & 0 & 0\\
0 & 0 & 0 & 1.1 & 0 & 0 & 0\\
0 & 0 & 0 & 0 & 1.1 & 0 & 0\\
\end{array}\right).
\end{equation}

\begin{figure}[h!]
	\begin{minipage}[h]{0.5\linewidth}
		\center{\includegraphics[width=1.1\linewidth]{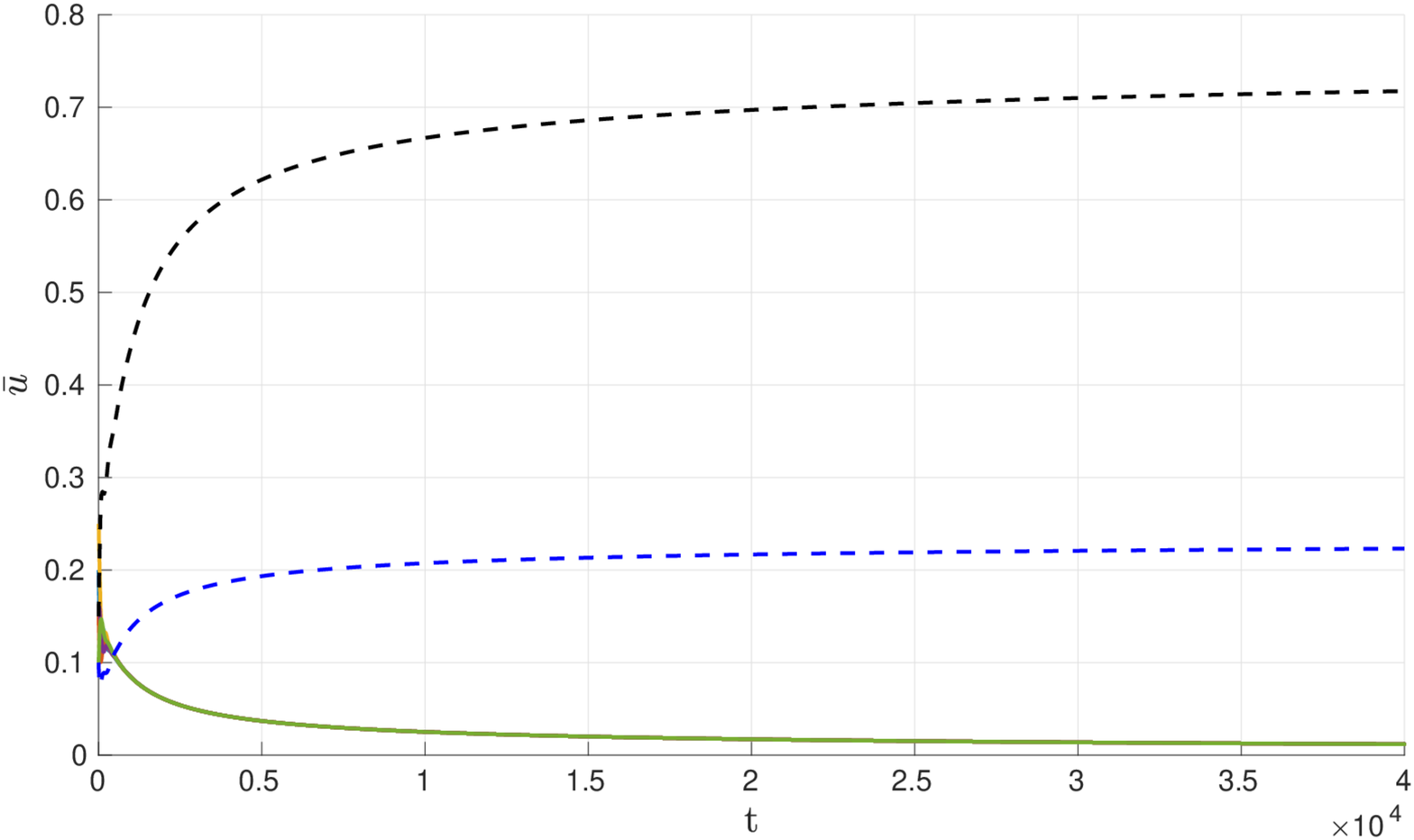}} a) \\
	\end{minipage}
	\hfill
	\begin{minipage}[h]{0.5\linewidth}
		\center{\includegraphics[width=1.1\linewidth]{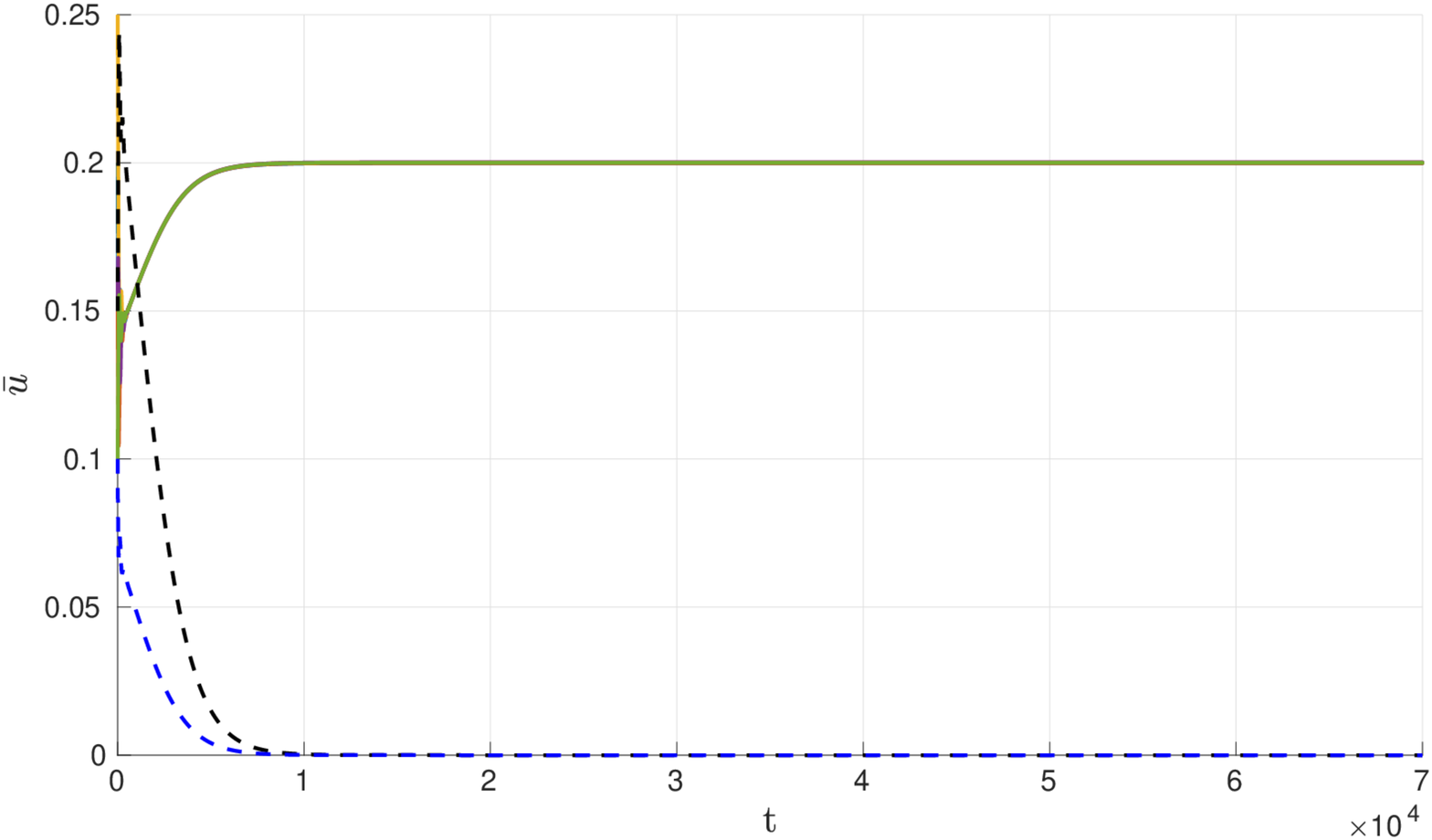}} \\b)
	\end{minipage}
	\vfill
	\begin{minipage}[h]{0.5\linewidth}
		\center{\includegraphics[width=1.1\linewidth]{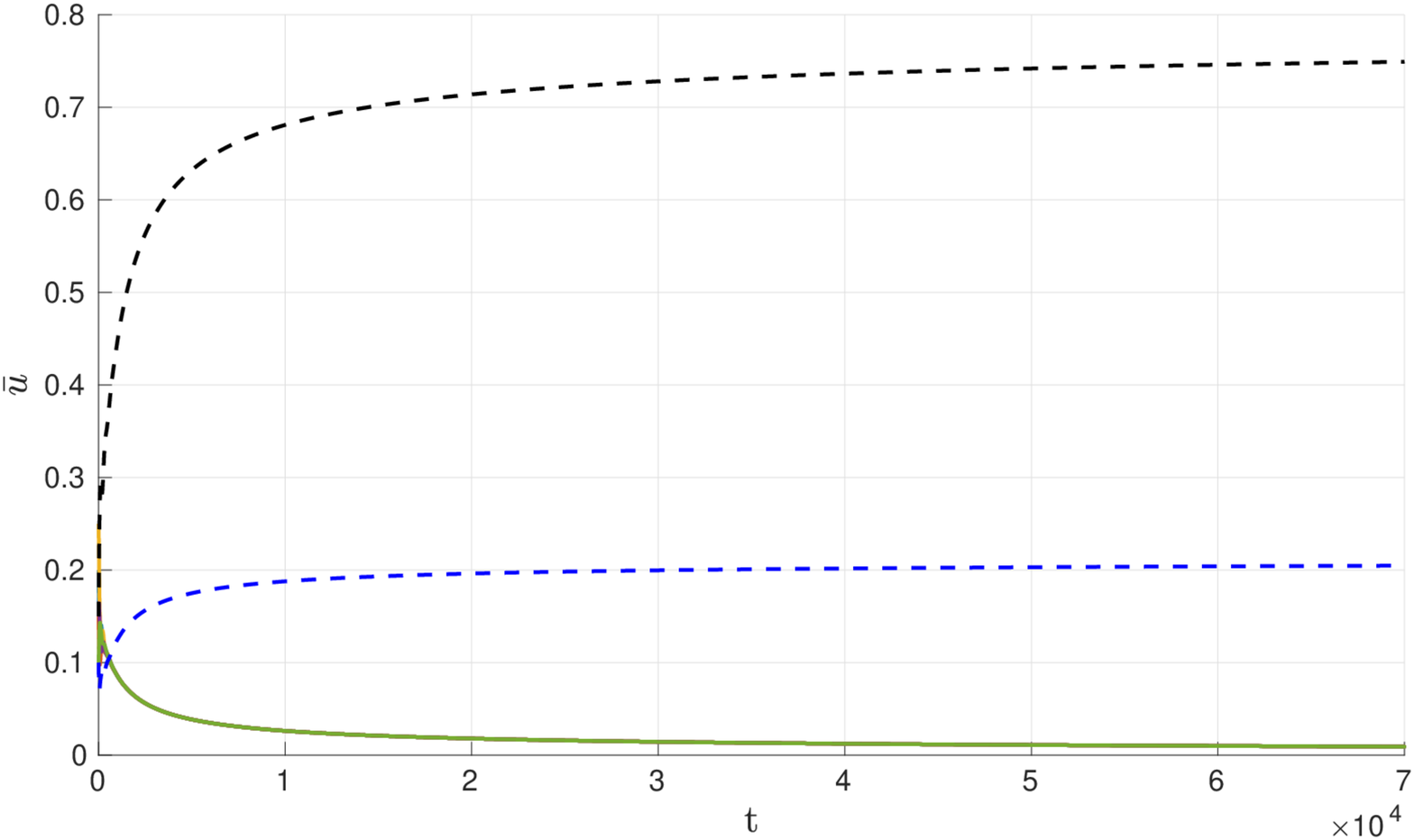}} c) \\
	\end{minipage}
	\hfill
	\begin{minipage}[h]{0.5\linewidth}
		\center{\includegraphics[width=1.1\linewidth]{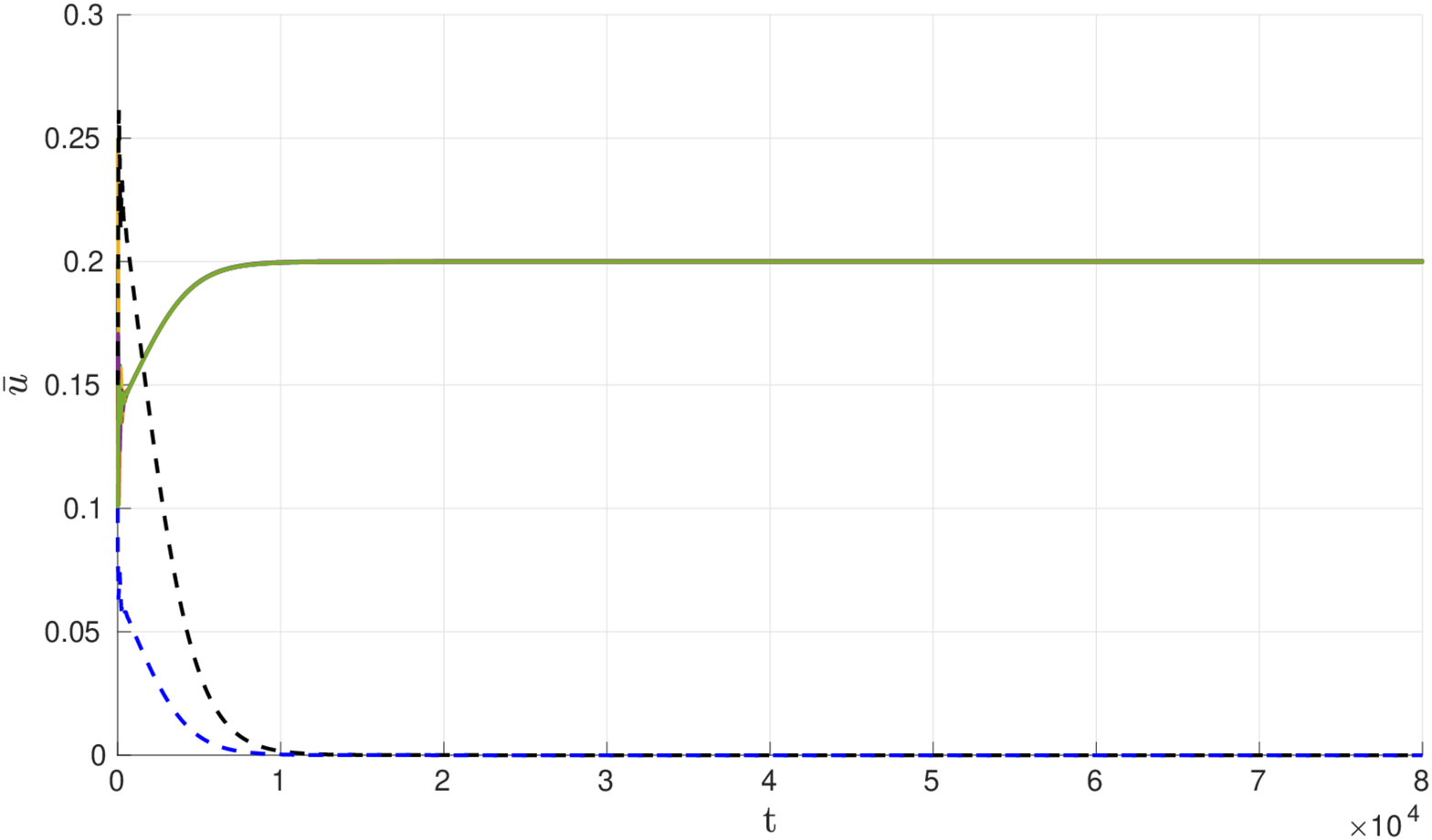}} d) \\
	\end{minipage}
	\caption{The frequencies of the species in the bi-hypercycle system with two parasites changing over system time $t$: a) with the matrix (\ref{e27}) at the beginning of the adaptation process; b) with the matrix (\ref{e28}) at the 30-th step of the fitness landscape evolutionary process; c)  with the matrix (\ref{e28}) and increased replication rate for parasites ($a_{64} = a_{75} = 1.2$) at the 30th step of the fitness landscape evolutionary process. d) with the increased replication rate for parasites ($a_{64} = a_{75} = 1.2$) at the 55-th step of the fitness landscape evolutionary process.}\label{fig6}
\end{figure}

As we have shown in a previous study \cite{drozh18}, evolved hypercycles obtain resistance to parasite invasion. The same property takes place with the bi-hypercycle system described above. Consider an example to illustrate this property. Fig. \ref{fig6}  shows how the cycle of length $n=5$ and its reaction to a parasite invasion. 
\newpage

\newpage
\section*{Evolution of the replicator system with new species adoptation at random time moments}
Consider the evolutionary process of the mean fitness value optimization of the replicator system (\ref{e1}, \ref{e2}). In \cite{drozh18},  we have shown an iteration process for hypercycles, leading to a steady growth of the function  $\bar{f}(\tau)$ over the matrix set under the condition  (\ref{e9}).

Let a new element of the replicator system appear at the random step of the iterative process $k$, which is characterized by $\tau_{k} = k \Delta\tau$.  
We denote a new matrix $(n + 1) \times (n + 1)$ as ${\bf A}^{k}$, which corresponds to the system (\ref{e1}, \ref{e2}) on  $k$-th step with $n\times n$ dimension. The steady-state distribution vector changes on this $k$ step to include the new species:  ${\bf \bar{u}}^{k} = (u_{1}^{k}, u_{2}^{k}, \ldots, u_{n + 1}^{k})$.

We assume the following:
\begin{enumerate}
	\item The mean fitness value in a steady-state of the system $\bar{f}(\tau_{k + 1})$ for $(k + 1)$-th step is the same as it should be without a new species:
	\begin{equation}\label{e29}
	\bar{f}(\tau_{k + 1}) = \bar{f}_{k + 1}. 
	\end{equation}
	
	\item Matrix elements ${\bf A}^{k + 1}$ for $(k + 1)$-th iteration do not change compare to  ${\bf A}^{k}$ for the first $n$ rows and $n$ columns, i.e. $a_{ij}^{k + 1} = a_{ij}^{k}, \quad i, j = 1,\ldots, n$.
	
	\item The first $n$ elements of the distribution vector for $k$-th and $(k + 1)$-th steps satisfy the condition:
	\begin{equation}\label{e30}
	\frac{u_{i}^{k + 1}}{u_{j}^{k + 1}} = \frac{u_{i}^{k}}{u_{j}^{k}}, \quad i, j = 1,\ldots, n.
	\end{equation}
 
	\item Impact of all species is the same, i.e.,
	
	\begin{equation}\label{e31}
	a_{n + 1, j}^{k + 1} u_{j}^{k + 1} = a_{n + 1, i}^{k + 1} u_{i}^{k + 1}, \quad i, j ={1,\ldots, n + 1}. 
	\end{equation}	
\end{enumerate}

From  (\ref{e30}), we obtain that a parameter $0 < \alpha_{k} < 1$ exists, such as the components of the vector ${\bf \bar{u}}^{k + 1}$ defined as follows:
\begin{equation}\label{e32}
u_{n+ 1}^{k + 1} = 1 - \alpha_{k} , \quad u_{i}^{k + 1} = \alpha_{k}u_{i}^{k}, \quad i, j = 1,\ldots, n. 
\end{equation}
It follows from (\ref{e32}) that:
$$
\sum\limits_{i = 1}^{n + 1} u_{i}^{k + 1} = \alpha_{k} \sum\limits_{i = 1}^{n} u_{i}^{k} + (1 - \alpha_{k}) = 1,
$$
$$
\bar{f}_{k + 1}  =\sum\limits_{j = 1}^{n + 1} a_{ij}^{k + 1}u_{j}^{k + 1} = \alpha_{k}\bar{f}_{k} + (1 - \alpha_{k})a_{i, n + 1}^{k + 1}.
$$
From the latter, we have:
\begin{equation}\label{e33}
	a_{i (n + 1)}^{k + 1} = \frac{\bar{f}_{k + 1} - \alpha_{k}\bar{f}_{k}}{(1 - \alpha_{k})}, \quad i, j = 1,\ldots, n. 
\end{equation} 
The formula (\ref{e33}) defines all the elements in $(n + 1)$-th column, besides the last element in the matrix  ${\bf A}^{k + 1}$.

The expression (\ref{e31}) gives the connection between  ${\bf A}^{k + 1}$:
\begin{equation}\label{e34}
\left\{
\begin{array}{rcl}

a_{(n + 1) j}^{k + 1} &=& \frac{\bar{f}_{k + 1}}{\alpha_{k}u_{j}^{k + 1}(n + 1)}, \quad j = 1,\ldots, n, \\
a_{(n + 1) (n + 1) }^{k + 1} &=& \frac{\bar{f}_{k + 1}}{(1 - \alpha_{k})(n + 1)}, \\

\end{array}
\right. 
\end{equation}
Thus, the procedure for the matrix  ${\bf A}^{k + 1}$ and the vector ${\bf \bar{u}}_{k + 1}$ at the time moment $k$ of the iteration process depends only on the parameter $0 < \alpha_{k} < 1$.

Numerical experiments have shown that if the frequency  $u_{n + 1}^{k + 1}$ is small enough for a new species, then $\alpha_{k} > \frac{1}{u_{\max}^{k} + 1}, \quad u_{\max}^{k + 1} = \max{u_{1}^{k}, \ldots, u_{n}^{k}}$, and the updated replicator system with  ${\bf A}^{k + 1}$ loses permanence. 
This means the violation of equality for the mean integral fitness and steady-state fitness value, which form the basis of the evolutionary adaptation process. 
If the frequencies $u_{n + 1}^{k + 1}$ of the new species are maximal, i.e., when  $\alpha_{k} < \frac{1}{u_{\max}^{k} + 1}$, $u_{\max}^{k + 1} = \max{u_{1}^{k}, \ldots, u_{n}^{k}}$, then the suggested evolutionary process of the mean fitness value maximization in a steady-state happens without losing the permanence property. 

The figures show the results of the numerical calculations for the evolutionary dynamics of the replicator system with new species adoption. The random value, which defines the number of step of the iteration process for the new species to enter the system, is described by the Poisson distribution. Using the uniform distribution, we define the parameter $\alpha_{k}$:
$$
\alpha_{k} = \beta\frac{1}{u_{\max}^{k} + 1}, \quad \beta \sim \mathbb{U}(0, 1).
$$ 
As the initial state  ($\tau = 0$), we consider (\ref{e26}). 

\begin{figure}[h!]
	\centering
	\includegraphics[scale=0.17]{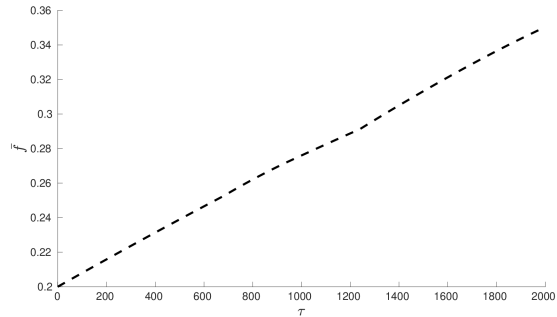}
	\caption{The mean fitness of the hypercycle system with two new elements changing over evolutionary time $\tau$ for the cycle length n = 5}
	\label{fig12}
\end{figure}

\begin{figure}[h!]
	\centering
	\includegraphics[scale=0.17]{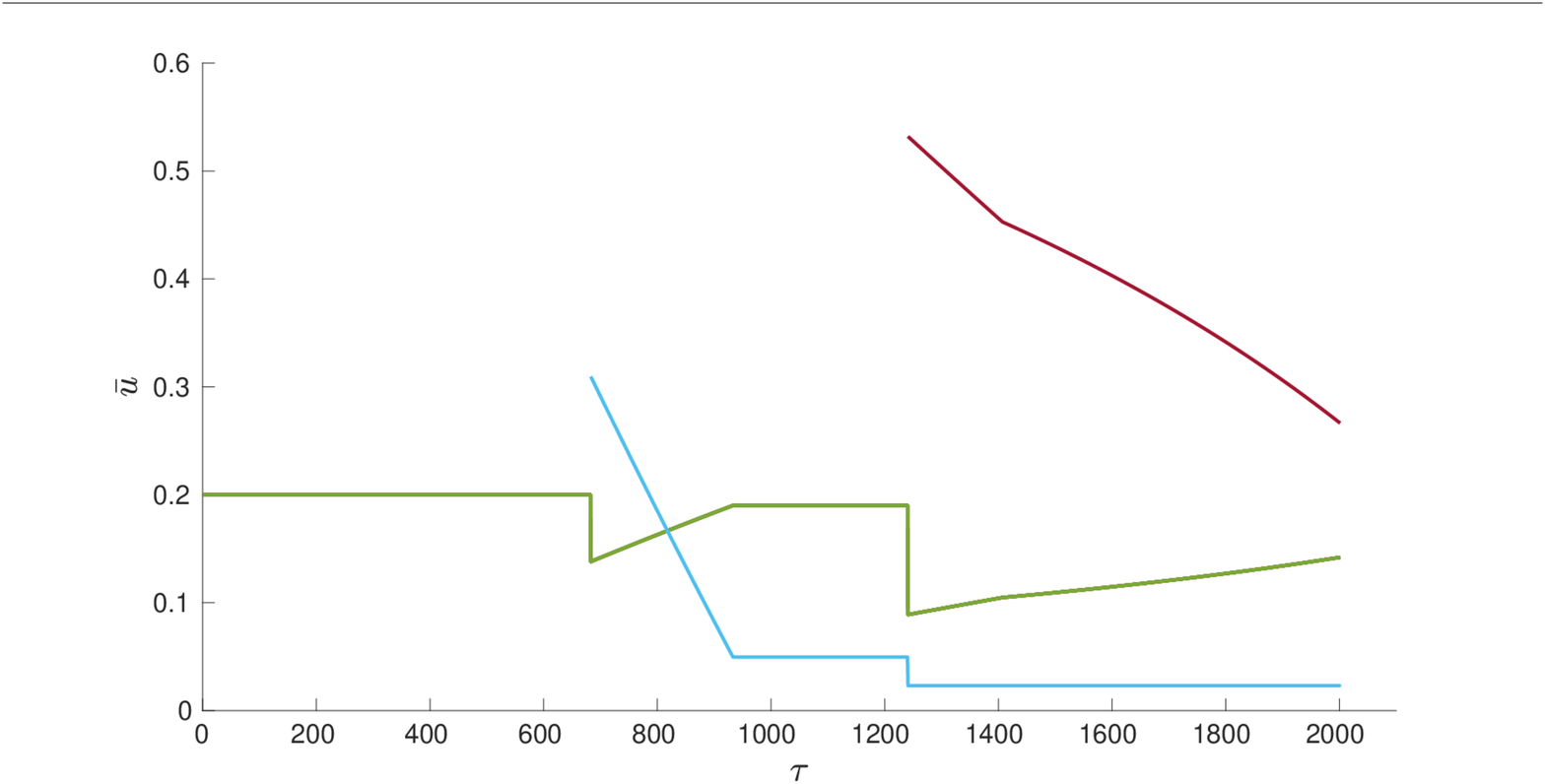}
	\caption{Steady---state ${\bf u} = \bar{u}$ of the hypercycle system with two new elements (2.1) changing over evolutionary time $\tau$ for the cycle length n = 5}
	\label{fig13}
\end{figure}

\begin{figure}[h!]
	\centering
	\includegraphics[scale=0.17]{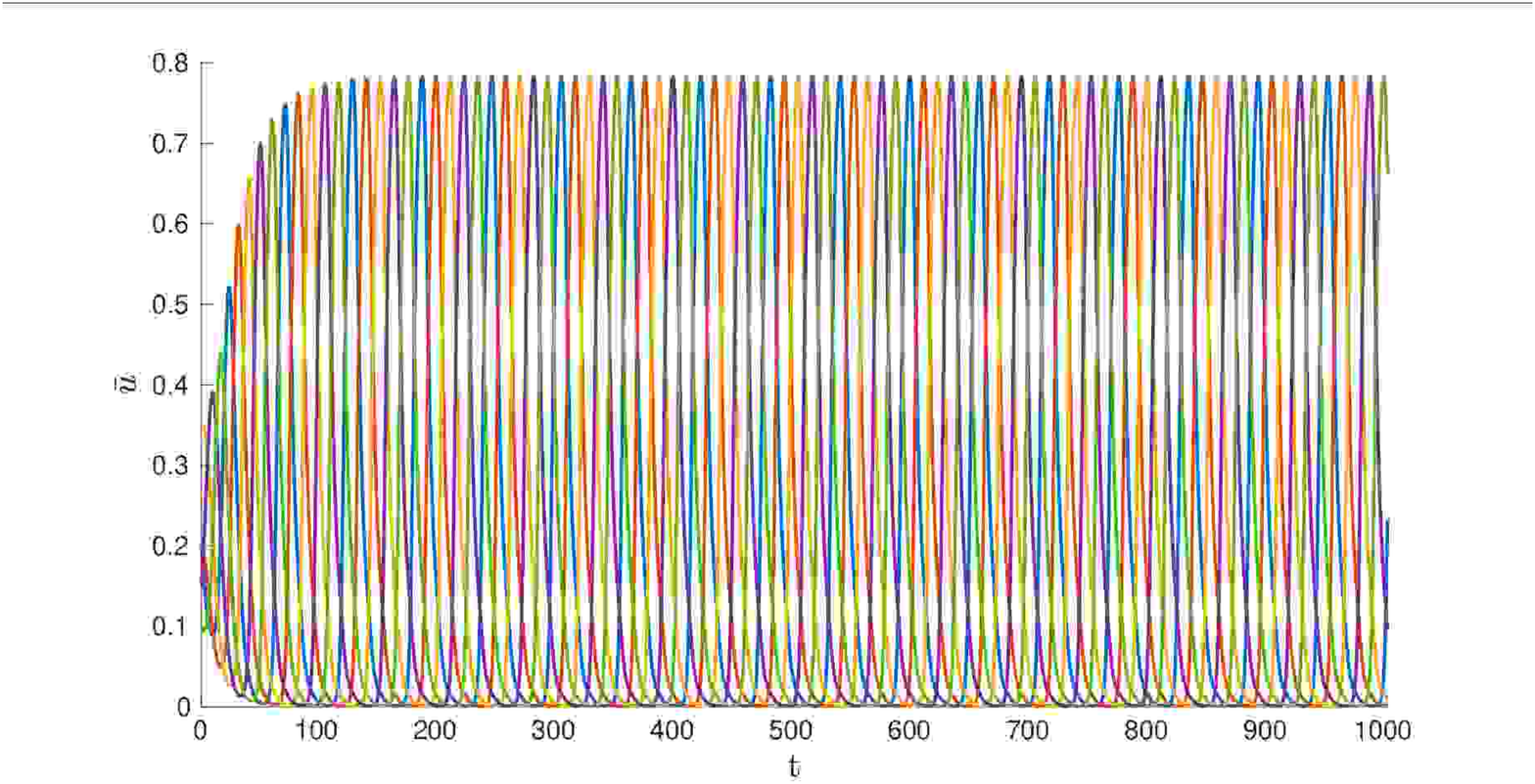}
	\caption{Frequencies of the species in the hypercycle system with two new elements (2.1) changing system time $t$ at the 681th step of the fitness landscape evolutionary process for the cycle length n = 5}
	\label{fig14}
\end{figure}

\begin{figure}[h!]
	\centering
	\includegraphics[scale=0.16]{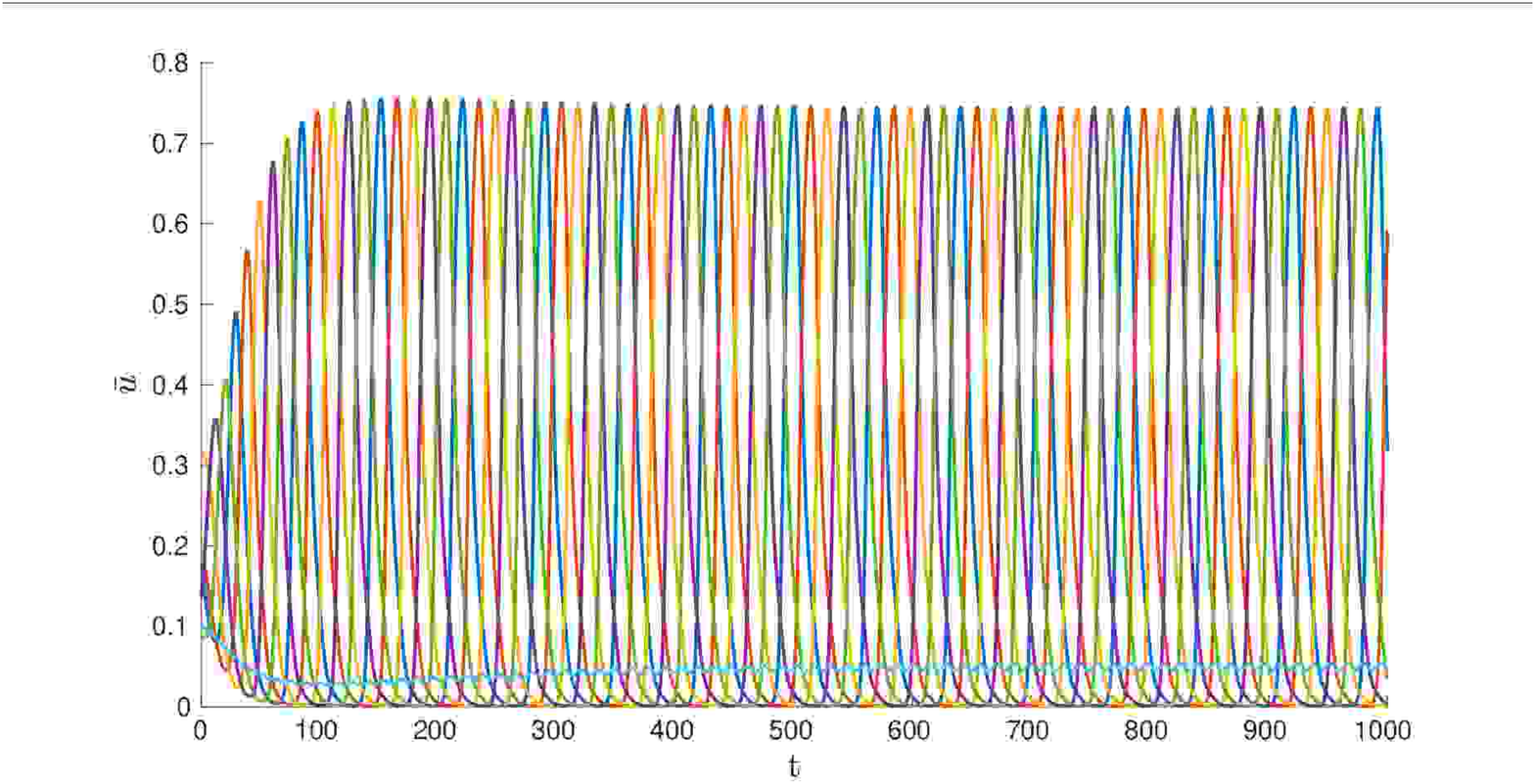}
	\caption{Frequencies of the species in the hypercycle system with two new elements (2.1) changing system time $t$ at the 1239th step of the fitness landscape evolutionary process for the cycle length n = 5}
	\label{fig15}
\end{figure}

\begin{figure}[h!]
	\centering
	\includegraphics[scale=0.16]{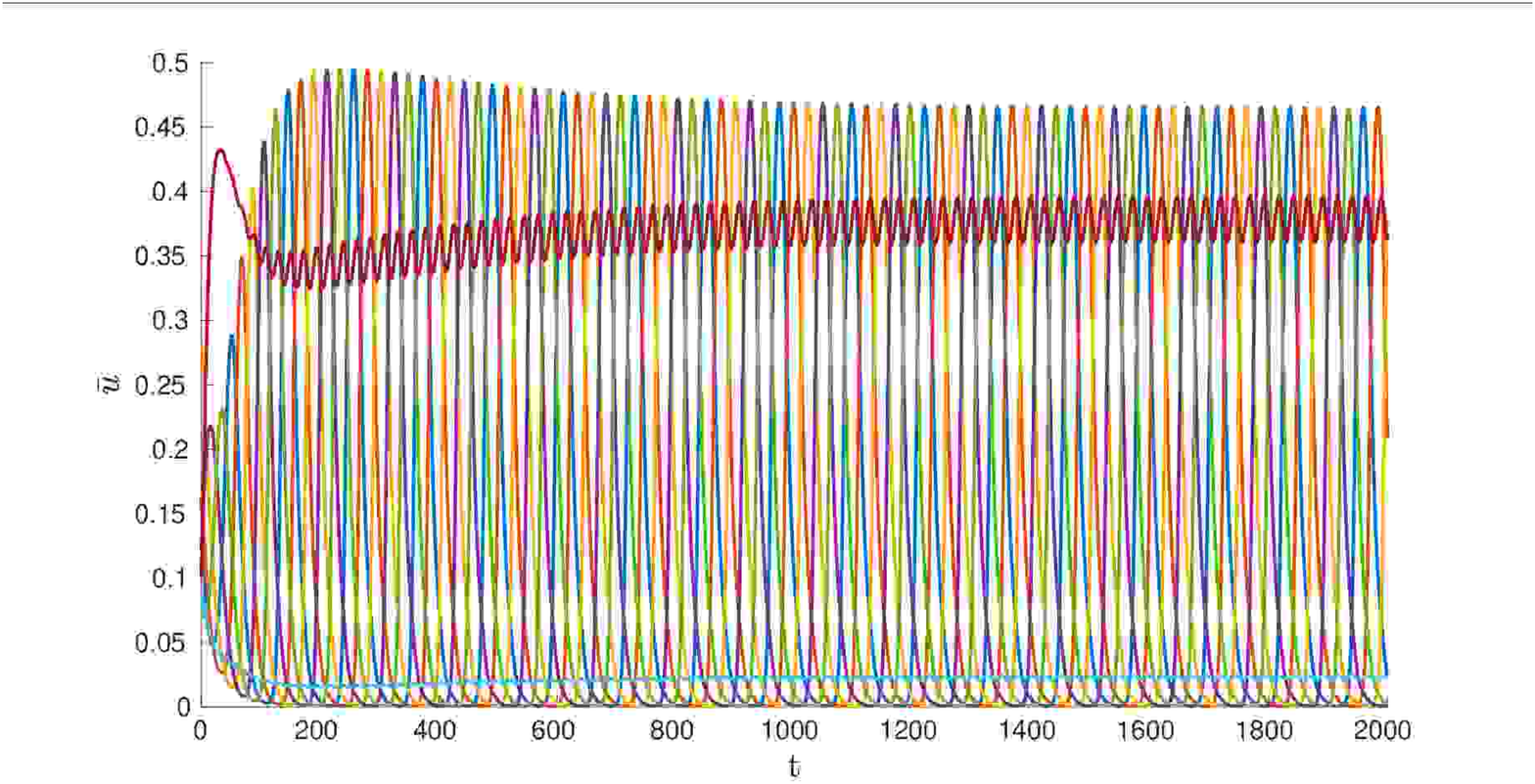}
	\caption{Frequencies of the species in the hypercycle system with two new elements (2.1) changing system time $t$ at the 1735th step of the fitness landscape evolutionary process for the cycle length n = 5}
	\label{fig16}
\end{figure}

We illustrate the of the evolutionary process with random species adoption in details below (based on the system (\ref{e1},\ref{e2}) for $n=5$):
\begin{itemize}
	\item[--] Figure \ref{fig12} shows the mean fitness in a steady-state. Here, the Poisson distribution adds new species at 682-th and 1240-th steps with the parameters $\beta_{682} = 0.829$ and $\beta_{1240} = 0.557$ correspondingly. 
	\item[--] Figure \ref{fig13} demonstrates the steady-state evolution. Green line defines the steady-state  for the first five elements of the system. Blue line corresponds to the steady-state of the system, which describes the sixth element: the first ``new'' one in the system. Red color used for the seventh element, which is the second new element after adding it to the system.
	\item[--] In Figure \ref{fig14}, we have the graphics for the frequency distribution in the replicator system over internal time $t$ before a new type entered the system.	
	\item[--] In Figure \ref{fig15}, the graphics describe the frequency distribution in the replicator system, depending on $t$ before the second new type was added to the system. Green line corresponds to the set, where the frequencies of the original 5 types fluctuates, blue --- for the sixth element. 	
	\item[--] Figure \ref{fig16} shows the frequency distribution in the replicator system, depending on $t$ at 1735-th iteration. As it was shown before, green color is used for the original 5 types, blue --- for the sixth and red --- for the seventh added at 1239-th step.  		
\end{itemize}

\begin{figure}[h!]
	\begin{minipage}[h]{0.47\linewidth}
		\center{\includegraphics[width=0.87\linewidth]{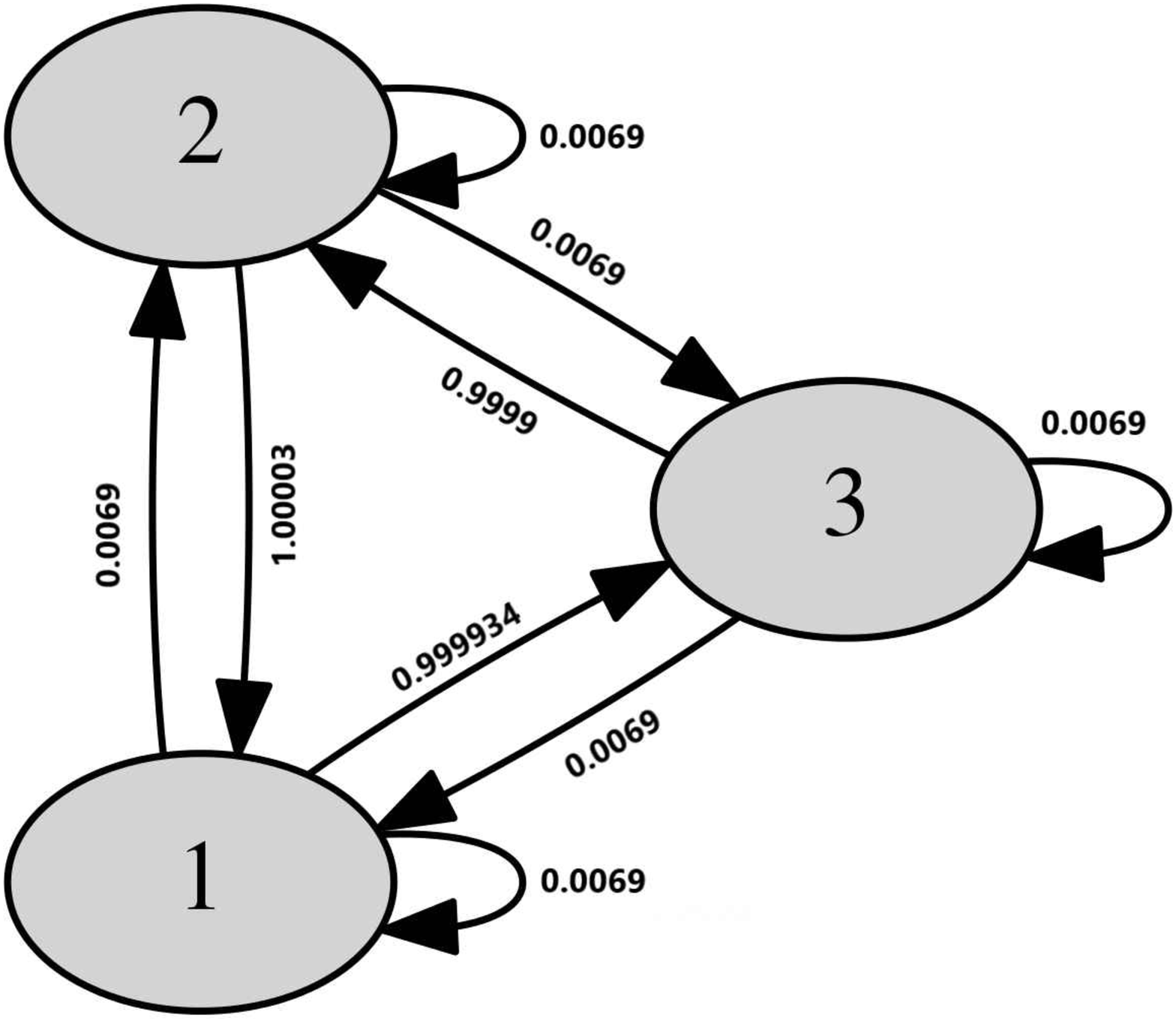} \\ a)}
	\end{minipage}
	\hfill
	\begin{minipage}[h]{0.47\linewidth}
		\center{\includegraphics[width=0.87\linewidth]{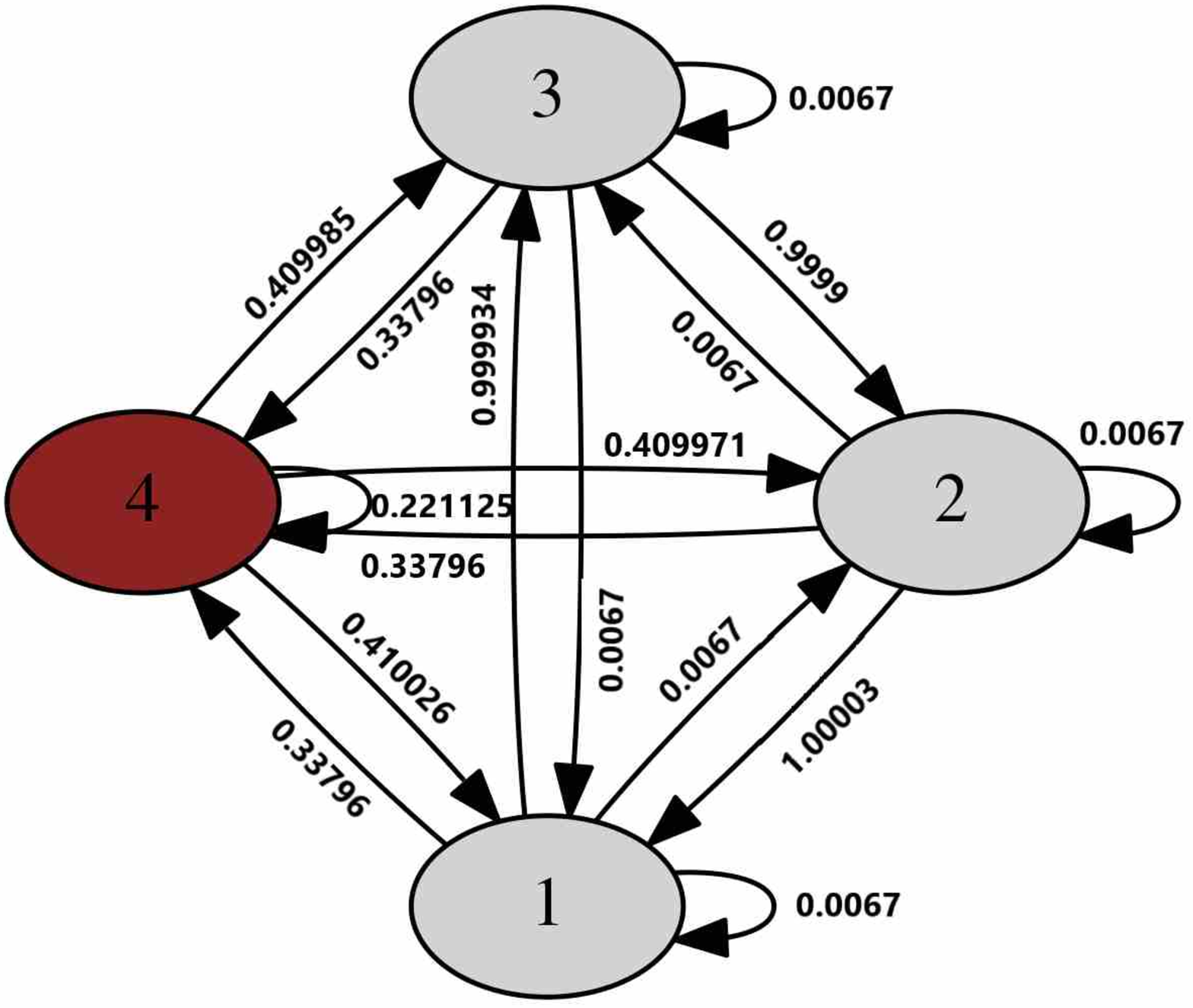} \\ b)}
	\end{minipage}
	\caption{a) Graph represents the evolved hypercycles: the result of the evolutionary process after 67 iterations. b) Graph represents the evolved hypercycles: the result of the evolutionary process after 68 iterations.}
	\label{fig17}
\end{figure}

\begin{figure}[h!]
	\begin{minipage}[h]{0.47\linewidth}
		\center{\includegraphics[width=0.87\linewidth]{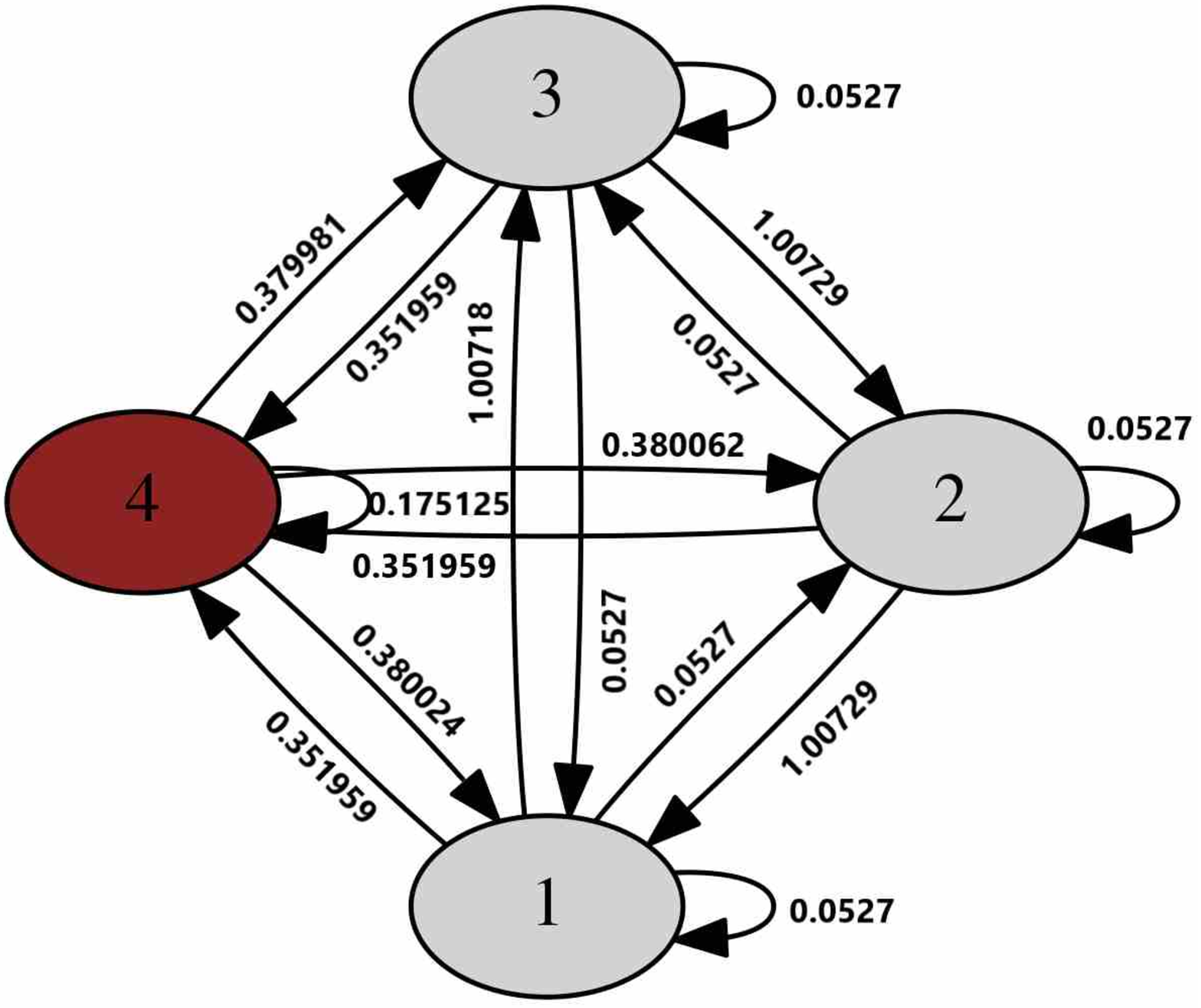} \\ a)}
	\end{minipage}
	\hfill
	\begin{minipage}[h]{0.47\linewidth}
		\center{\includegraphics[width=0.87\linewidth]{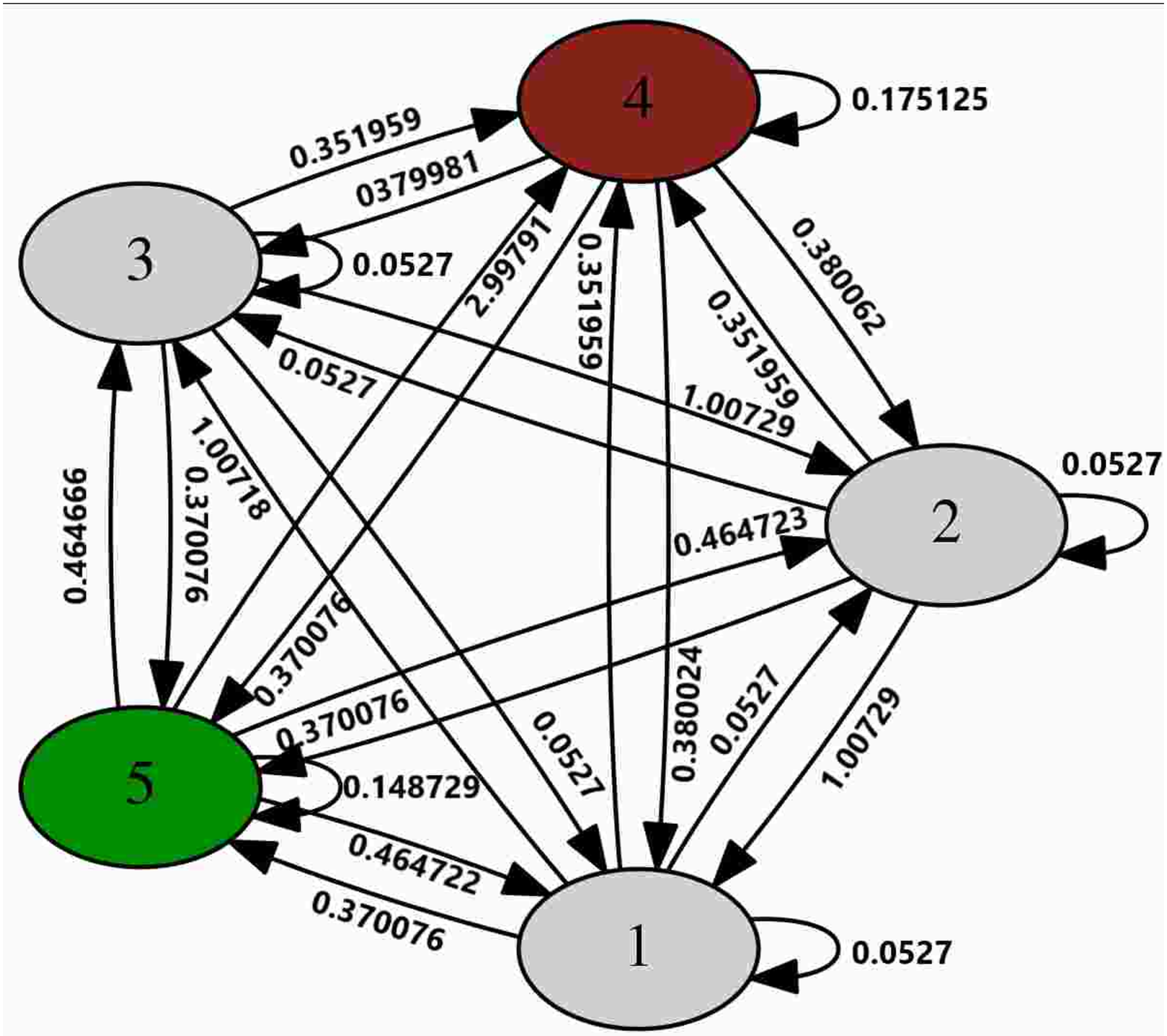} \\ b)}
	\end{minipage}
	\caption{a) Graph represents the evolved hypercycles: the result of the evolutionary process after 528 iterations. b) Graph represents the evolved hypercycles: the result of the evolutionary process after 529 iterations.}
	\label{fig18}
\end{figure}

To show, how the hypercycle systems evolve under this assumption of new species inclusion, consider a third-order system. In Figure \ref{fig17} (a-b), we see how the initial state of the system transforms after the fourth type appearance. Figure \ref{fig18}(a-b) depicts the state before the fifth type and the transitions after the further extension. These results show the possibility of a significant evolutionary change of the system, which is extended by a new type at random time moments.


\newpage
\section*{Special settings for replicator systems}
\subsection*{``Ant hill'' system}
\begin{figure}[h!]
	\centering
	\includegraphics[width=0.85 \textwidth]{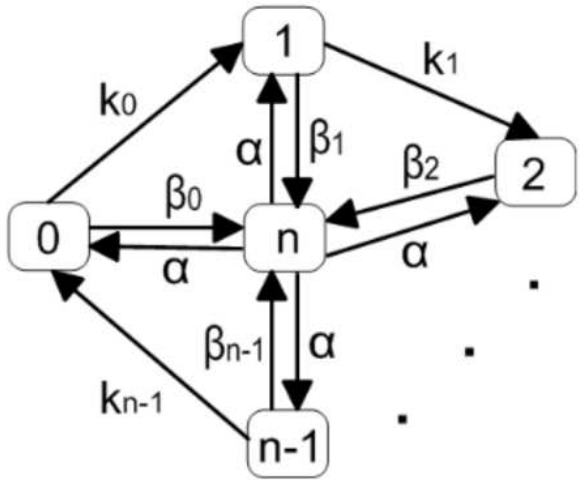}
	\caption{The graph represents the ``ant hill''\, system}
	\label{fig19}
\end{figure}
Consider a system with transitions defined by  Figure \ref{fig19}. Using the association with biological populations dynamics, we call this class of systems ``ant hill''. In this case, we have a dominating type which is supported by  $n - 1$ other types. Here,  catalyzic coefficients are  $\beta_{i},  i = 1, \ldots, n - 1,$ for the dominating type and all the backwards coefficients are $\alpha$. Moreover, there is cycle for general types with coefficients $k_{i}$.

The system state is described by the following:
\begin{equation}\label{e35}
\left\{
\begin{array}{rcl}

\dot{u}_{i} &=& u_{i}(\alpha u_{n} + k_{i}u_{i - 1} - f(t)), \quad i = {0,\ldots, n - 1},\\
\dot{u}_{n} &=& u_{n}\Bigg(\sum\limits_{i = 0}^{n - 1}\beta_{i} u_{i} - f(t)\Bigg), \\
f(t) &=& \alpha u_n\sum\limits_{i = 0}^{n - 1}u_{i} + \sum\limits_{i = 0}^{n - 1}k_{i}u_{i}u_{i - 1} + u_{n}\sum\limits_{i = 0}^{n - 1}\beta_{i}u_{i},\\

\end{array}
\right.
\end{equation}
where $ u_{-1} = u_{n - 1}, \quad u_{i}(0) = u_{i}^{0},\quad {\bf \bar{u}} \in S_{n}, \quad \alpha > 0, \quad \beta_{i} > 0, \quad k_{i} > 0, \quad i = {0, \ldots,n - 1}.$
It is natural to consider $n\geqslant3$ The system (\ref{e35}) has a unique steady-state $\bar{\bfu} \in intS_{n+1},$ which is a necessary condition for the system to be permanent. 

From (\ref{e35}) it follows: 
\begin{equation}\label{e36}
	k_i\bar{u}_{i-1} = f - \alpha_{n}\bar u_n, \quad i =0,\ldots, n-1.
\end{equation}
Hence $k_1\bar{u}_0 = k_2\bar{u}_1 = \ldots = k_{n-1}\bar{u}_{n-2} = k_{0}\bar{u}_{n-1}.$
$$
\bar{u}_i = \frac{k_1}{k_{i+1}}\bar{u}_0, \quad k_n = k_0,  i = 1,\ldots, n-1,.
$$
Thus, 
$$
\alpha_{n}\bar{u}_n+k_1\bar{u}_0 = \beta_0\bar{u}_0 + \bar{u}_0\sum_{j=1}^{n-1}\beta_j\frac{k_1}{k_{j+1}}, k_n=k_0.
$$
Since $\bar{u}_n = 1-\sum_{j=0}^{n-1}\bar{u}_j$, then 
$$
\bar{u}_0 = \alpha\left[\left((\alpha+1)\sum_{j=1}^{n-1}\frac{1}{k_{j+1}}-1\right)k_1+\alpha +\beta_0\right]^{-1}>0
$$
Let us prove, that the sufficient condition for the  (\ref{e35}) system to be permanent consists of several conditions on the system's parameters. 

Introduce the notation:
$$
k_{M} = \max{(k_0, k_{1}, \ldots, k_{n - 1})}, \quad k_{m} = \min{(k_0, k_{1}, \ldots, k_{n - 1})}, $$
$$
\beta_{M} = \max{(\beta_{0}, \beta_{1}, \ldots, \beta_{n - 1}}), \quad
\beta_{m} = \min{(\beta_{0}, \beta_{1}, \ldots, \beta_{n - 1})}.
$$

\begin{thm} 
Let the condition hold:
	\begin{equation}\label{e37}
	\beta_{m} > k_{M}, \alpha +\beta_m > \frac{k_m}{n} > \beta_{M}, n=3,4,\ldots, N.
	\end{equation}
If such values of the parameters exist, then (\ref{e37}) is permanent.
\end{thm}	
Consider the function: 
$$
\Phi(\bfu) = \ln \prod_{i=0}^{n-1}(u_i(t))^{\frac{1}{n}} - \ln u_n(t).
$$
Then 
$$
\dot{\Phi}(\bfu) = \frac{1}{n}\sum_{i=0}^{n-1}\frac{\dot{u}_i}{u_i} - \frac{\dot{u}_n}{u_n} = 
\alpha u_n + n\sum_{i=0}^{n-1}k_iu_{i-1} - n\sum_{i=0}^{n-1}\beta_iu_i.
$$
The estimation works
\begin{eqnarray}\label{e38}
\sum_{i=0}^{n-1}k_iu_{i-1} \geqslant k_m\sum_{i=1}^{n-1}u_i = k_m(1-u_n),\nonumber\\
\sum_{i=0}^{n-1}\beta_iu_{i} \leqslant \beta_M\sum_{i=1}^{n-1}u_i = \beta_M(1-u_n).
\end{eqnarray}
Taking into account (\ref{e37}), we derive:
$$
\dot{\Phi}(\bfu) \geqslant \delta_0 >0,\quad \delta_0 = \frac{k_m}{n}-\beta_{M}>0.
$$ 
Hence, 
\begin{equation}\label{e39}
	\prod_{i=0}^{n-1}(u_i(t))^{\frac{1}{n}}\geqslant c e^{\delta t}u_n(t).
\end{equation}
Consider a function $S(t) = \sum_{i=0}^{n}u_i(t)$:
$$
\dot{S}(t) = \alpha S(1-S) - \alpha(1-S)S^2 + (1-S)\sum_{i=0}^{n-1}k_iu_iu_{i-1} - S(1-S)\sum_{i=0}^{n-1}\beta_{i}u_i.
$$
From (\ref{e38}) and 
$$
\sum_{i=0}^{n-1}k_iu_iu_{i-1} \leqslant k_MS, 
$$
we get:
$$
\dot{S}(t) \leqslant (\alpha+\beta_m)S(S-1)(S-r^2).
$$
Here $r^2 = \frac{k_{M+\alpha}}{\beta_{m+\alpha}} <1.$ From the comparison theorem, we get:
$$
S(t) \leqslant \max\{r^2, \phi^2\}.
$$
Where $\phi^2 = S(0) = \sum_{i=0}^{n-1}u_i(0)<1$, $u_n(0) = 1-\phi^2 >\varepsilon_o >0.$
Hence, 
$$
u_n(t) = 1-S(t) \geqslant \min \{1-r^2, 1-\phi^2\}, 
$$
which completes the proof. 
\\ $\Box$

To illustrate this analysis, we take $n = 5$, where:
\begin{equation}\label{e40}
{\bf A} = \left(
\begin{array}{ccccc}
0 & 0 & 0 & 1 & 0.1\\
1 & 0 & 0 & 0 & 0.1\\
0 & 1 & 0 & 0 & 0.1\\
0 & 0 & 1 & 0 & 0.1\\
0.8 & 0.8 & 0.8 & 0.8 & 0\\
\end{array}\right),
\end{equation}
and $\alpha = 0.1, \quad \beta_{i} = 0.8, \quad k_{i} = 1$.

\begin{figure}[h!]
	\begin{minipage}[h]{0.4\linewidth}
		\center{\includegraphics[width=0.87\linewidth]{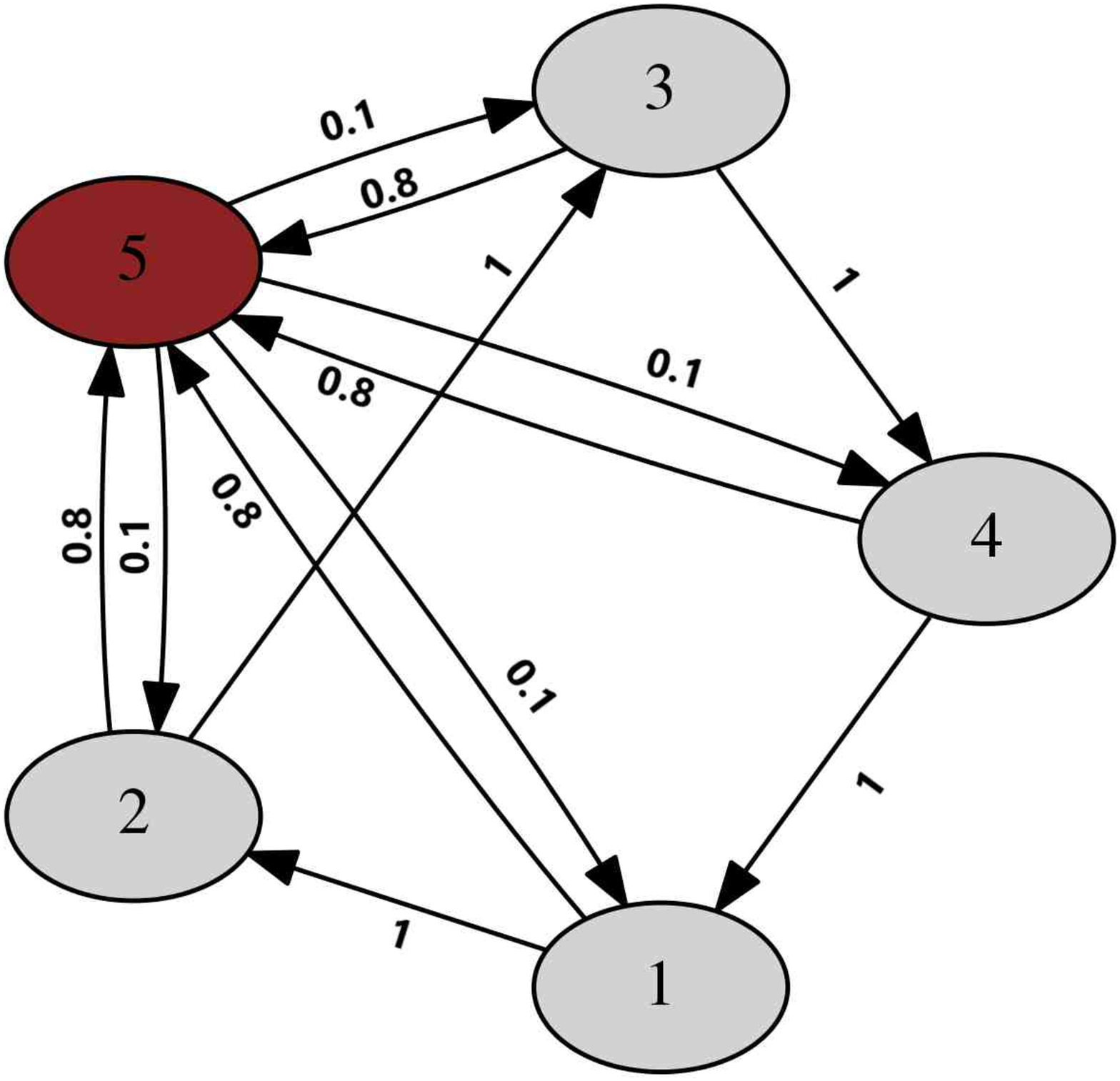} \\ a)}
	\end{minipage}
	\hfill
	\begin{minipage}[h]{0.4\linewidth}
		\center{\includegraphics[width=0.87\linewidth]{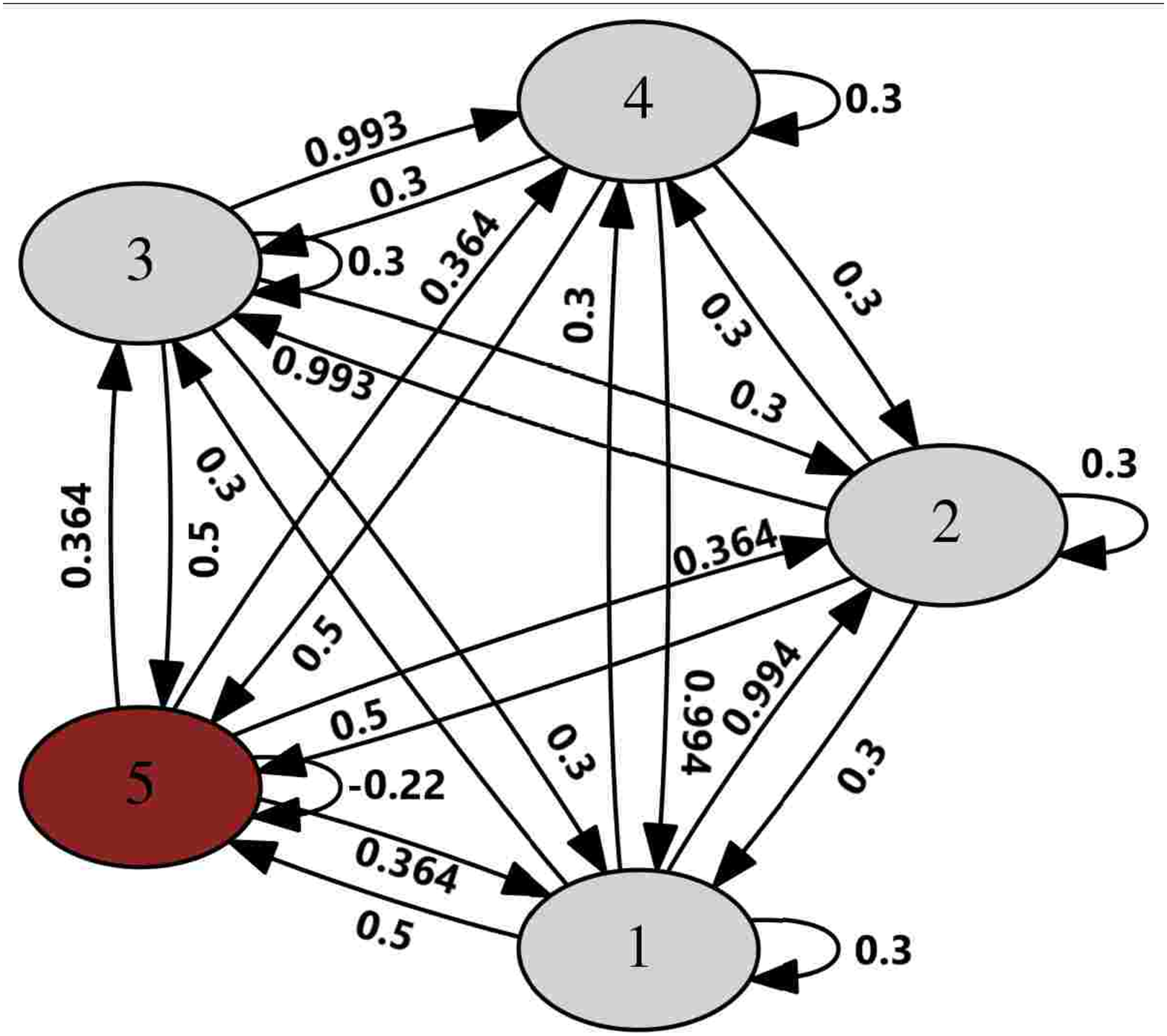} \\ b)}
	\end{minipage}
	\caption{ Graph represents the original hypercycles with matrix {\bf A} (\ref{e40}): a) before the evolutionary process.  b) the result of the evolutionary process after 300 iterations. Numbers on the edges correspond to the adapted fitness coefficients}
	\label{fig21}
\end{figure}

Evolutionary changes during the system's adaptation and mean fitness maximization lead to 
reduction of the parameters $\beta_{i}$, which define the catalysis of the dominating macromolecule  (Fig. \ref{fig21}). The corresponding steady-state describing dominating macromolecule converges to zero (approximately at 310-th step).

\begin{figure}[h!]
	\begin{minipage}[h]{0.5\linewidth}
		\center{\includegraphics[width=1.1\linewidth]{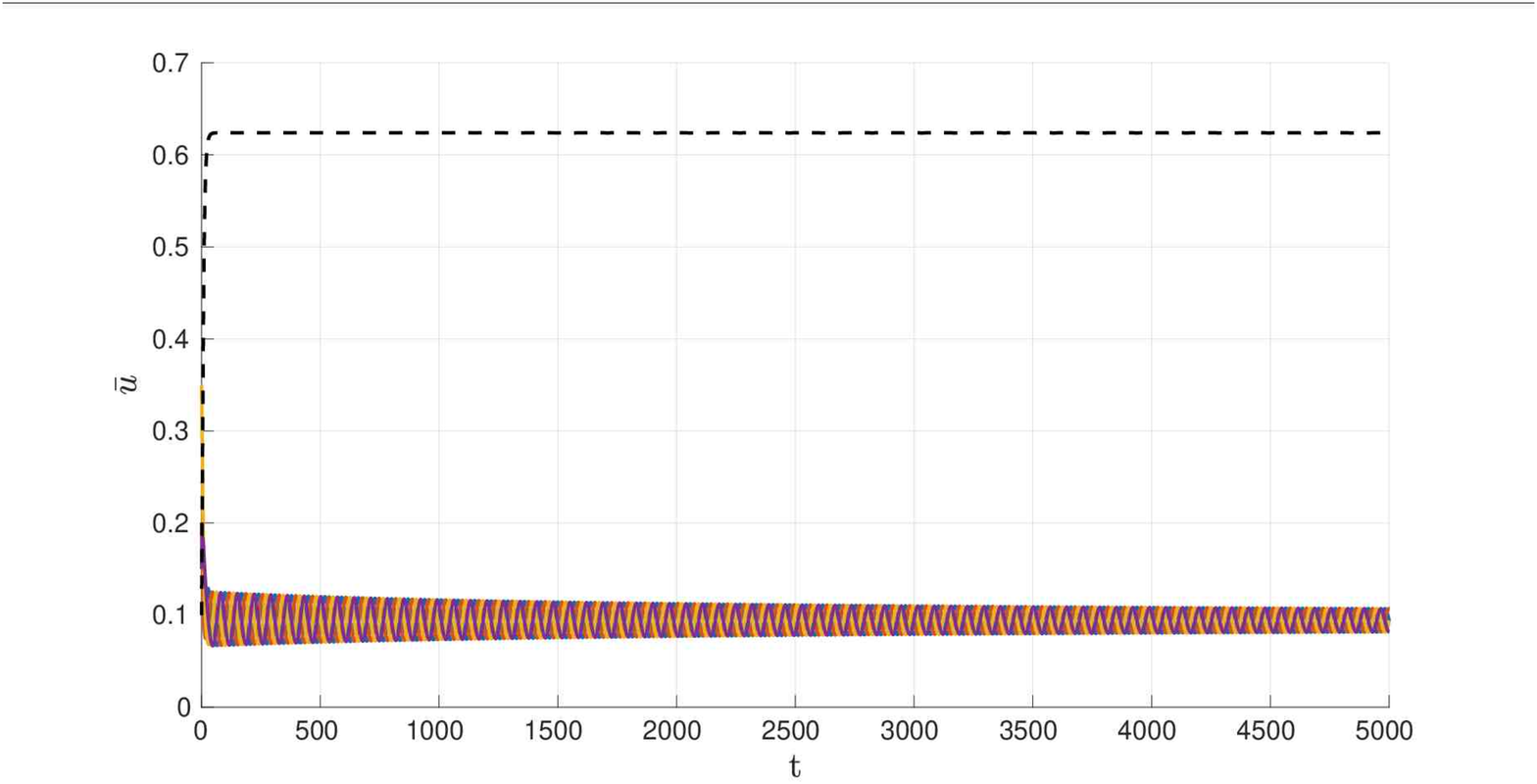}} a) \\
	\end{minipage}
	\hfill
	\begin{minipage}[h]{0.5\linewidth}
		\center{\includegraphics[width=1.1\linewidth]{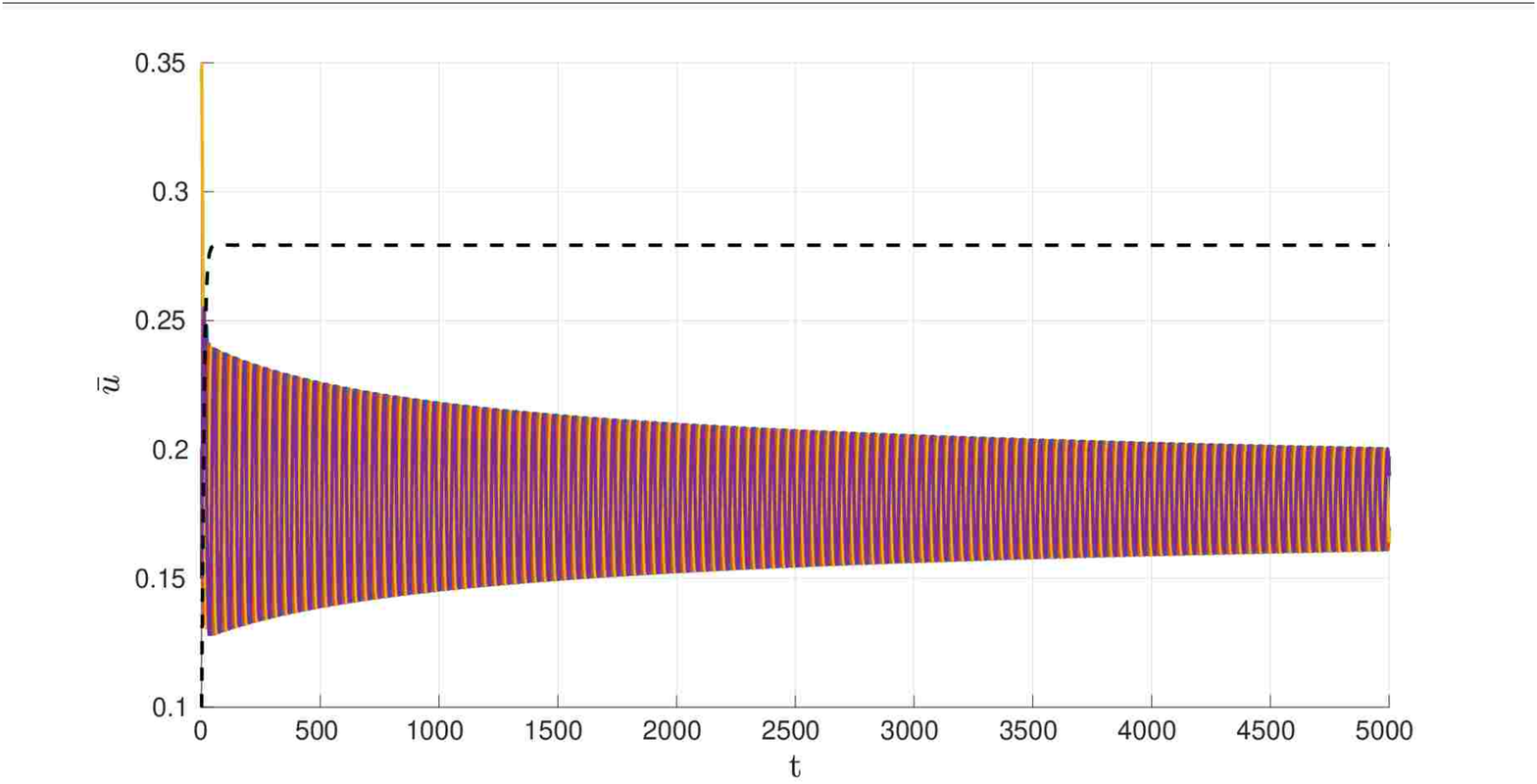}} \\b)
	\end{minipage}
	\vfill
	\begin{minipage}[h]{0.5\linewidth}
		\center{\includegraphics[width=1.1\linewidth]{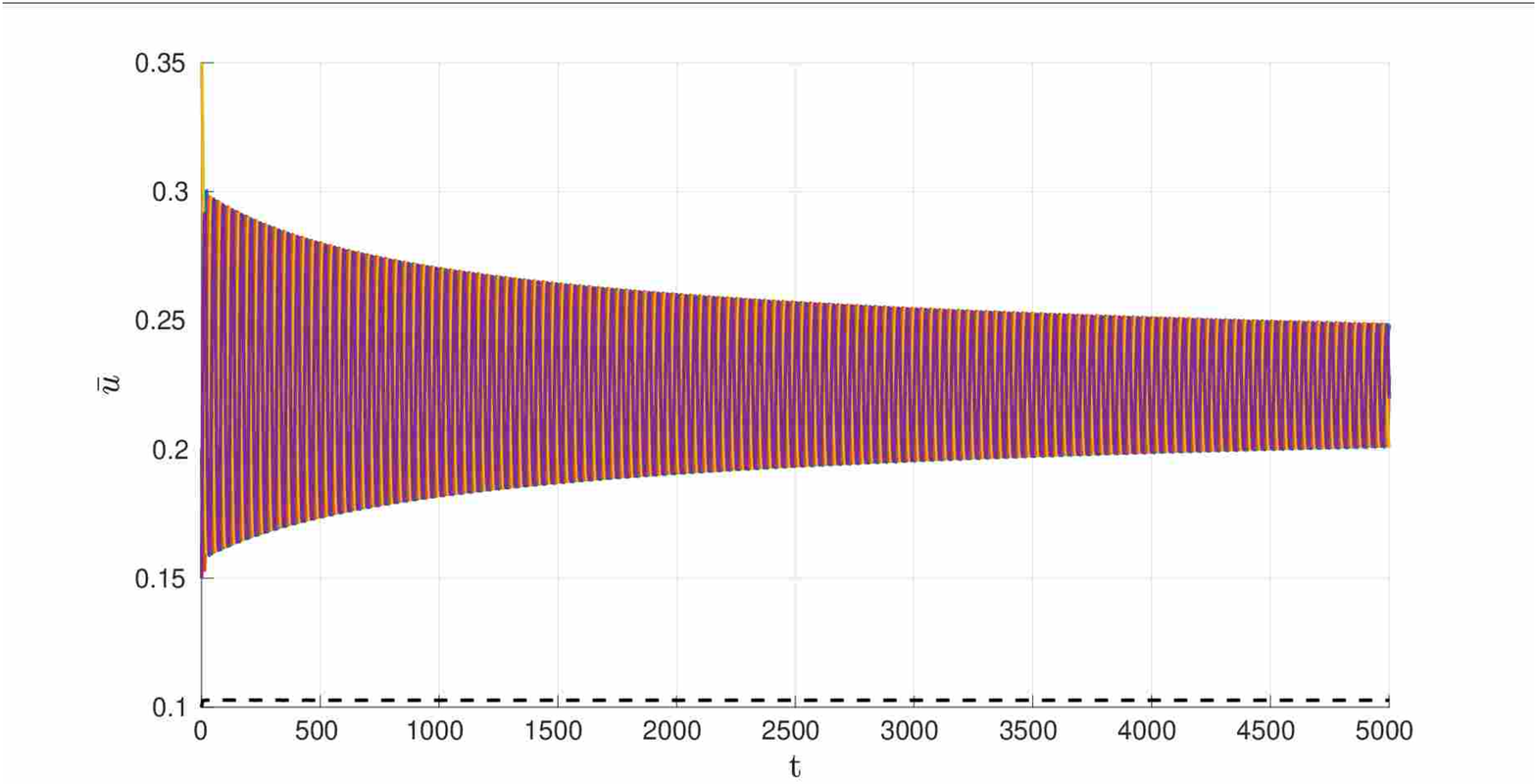}} c) \\
	\end{minipage}
	\hfill
	\begin{minipage}[h]{0.5\linewidth}
		\center{\includegraphics[width=1.1\linewidth]{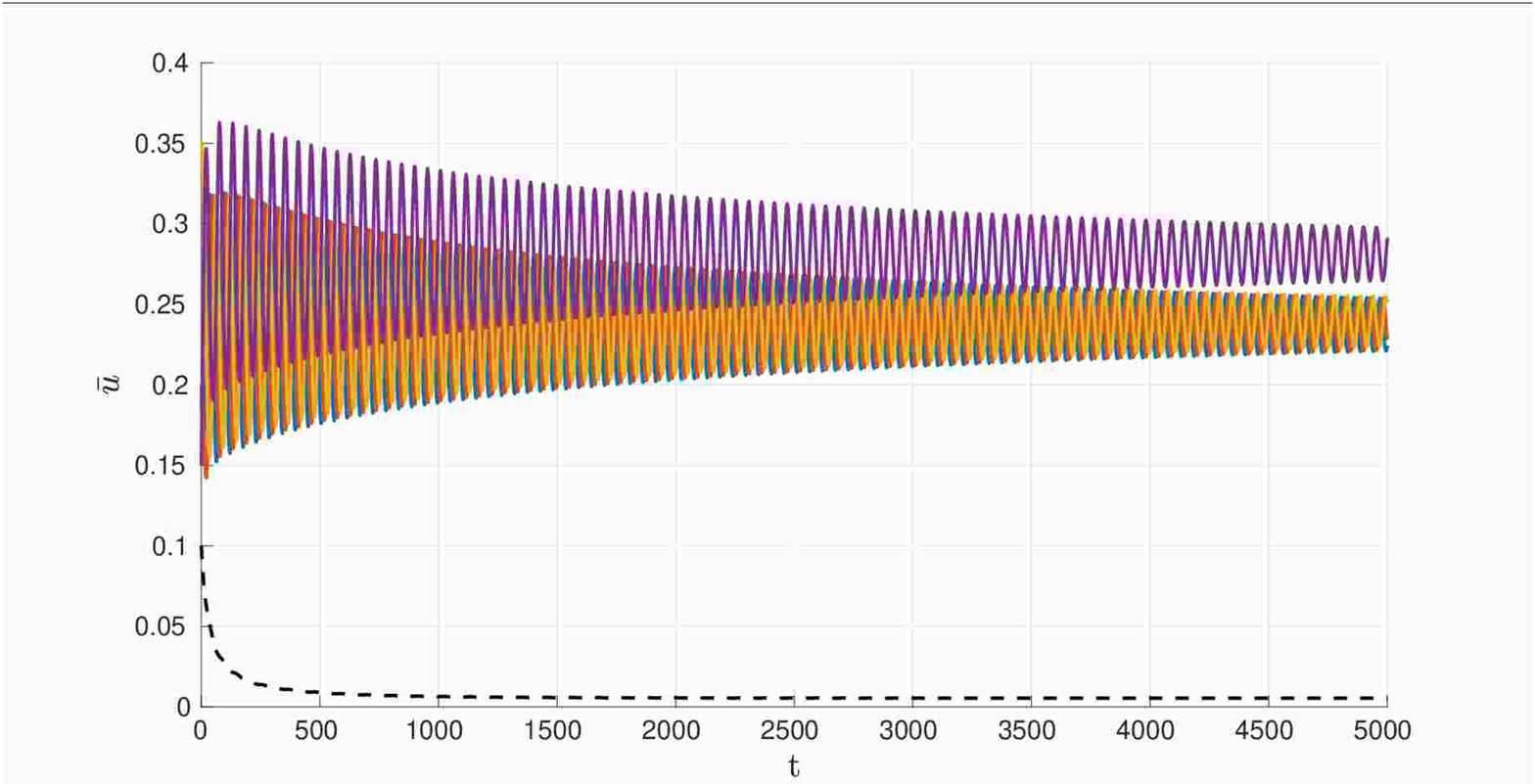}} d) \\
	\end{minipage}
	\caption{The frequencies of the species in the hypercycle system  with matrix {\bf A} (\ref{e40}) changing over system time $t$ at the $N$-th step of the fitness landscape evolutionary process. a) $N=50$ b) $N =175$ c) $N=250$ d) $N=400$.}
	\label{fig20}
\end{figure}

Figure \ref{fig20} shows dominating molecule (dotted line) and general molecules (solid line) at 50, 175, 250, and 400 steps of the evolution of the system (\ref{e35}) with ${\bf A}$ (\ref{e40}). It is worth mentioning, that the influence of the dominating molecule goes down and the amplitude of all others increase. 
\newpage

\newpage
\subsection*{Hypercycle with two types of behaviour}
Consider the system that describes one of the experimental models for the  RNA  cooperative networks \cite{rna12}, which is described in Figure \ref{fig23} :
\begin{figure}[h!]
	\begin{minipage}[h]{0.47\linewidth}
		\center{\includegraphics[width=0.87\linewidth]{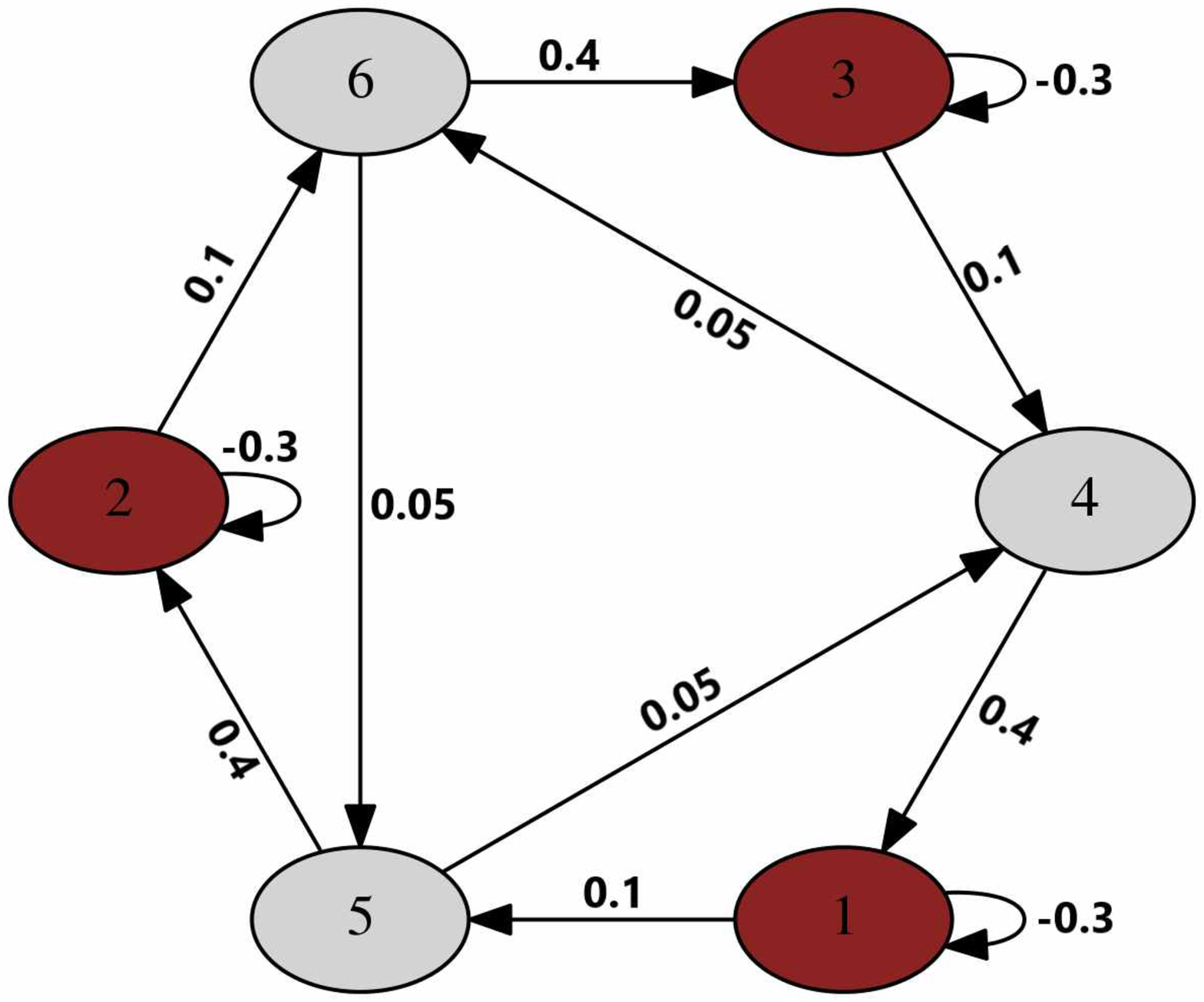} \\ a)}
	\end{minipage}
	\hfill
	\begin{minipage}[h]{0.47\linewidth}
		\center{\includegraphics[width=0.87\linewidth]{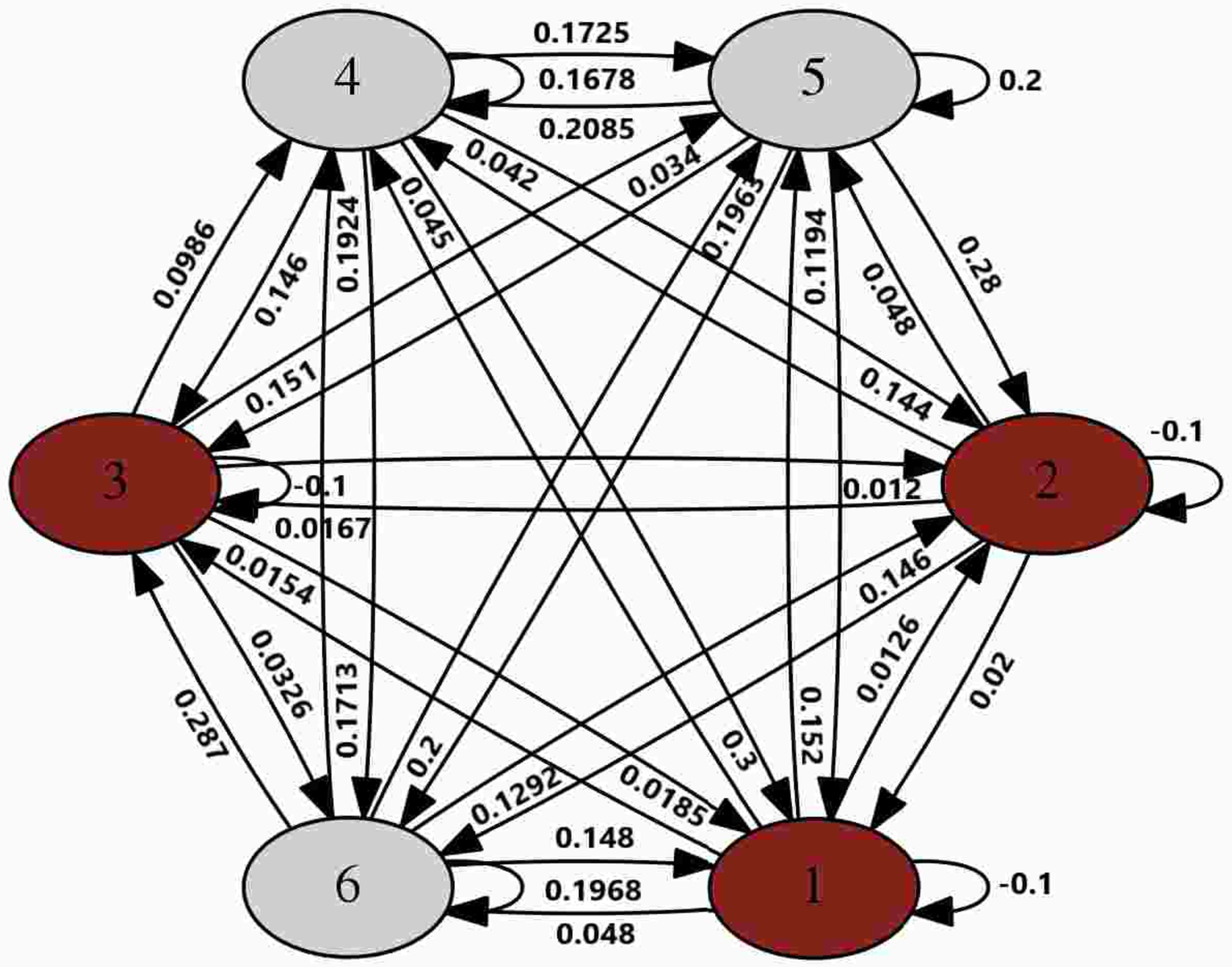} \\ b)}
	\end{minipage}
	\caption{Graph represents the original hypercycles (\ref{e41}) with matrix {\bf A} (\ref{e42}): a) before the evolutionary process, b)  the result of the evolutionary process after 200 iterations. Numbers on the edges correspond to the adapted fitness coefficients}
	\label{fig23}
\end{figure}
\begin{eqnarray}\label{e41}
\dot{u}_{1} &=& u_{1}(\alpha u_{1} + \beta u_{4} - f), \nonumber\\
\dot{u}_{2} &=& u_{2}(\alpha u_{2} + \beta u_{5} - f), \nonumber\\
\dot{u}_{3} &=& u_{3}(\alpha u_{3} + \beta u_{6} - f), \nonumber\\
\dot{u}_{4} &=& u_{4}(\sigma u_{3} + \gamma u_{5} - f), \\
\dot{u}_{5} &=& u_{5}(\sigma u_{1} + \gamma u_{6} - f), \nonumber\\
\dot{u}_{6} &=& u_{6}(\sigma u_{2} + \gamma u_{4} - f). \nonumber
\end{eqnarray}
Here, 
$$
f = \Big({\bf A\bar{u}, \bar{u}}\Big) = u_{1}(\alpha u_{1} + \beta u_{4}) + u_{2}(\alpha u_{2} + \beta u_{5}) + u_{3}(\alpha u_{3} + \beta u_{6}) + u_{4}(\sigma u_{3} + \gamma u_{5}) +
$$
$$ + u_{5}(\sigma u_{1} + \gamma u_{6}) + u_{6}(\sigma u_{2} + \gamma u_{4}).
$$
This system describes a population, which consists of two types --- ``egoists'' and  ``altruists''. We call egoists the molecules 1, 2, and 3, which are participating  in autocatalysis  with coefficient $\alpha$, and one of the 4, 5, or 6 with coefficient $\sigma$. Altruists, in this case, are 4, 5, and 6, which enforce the catalysis of others: egoists --- with  $\beta$and one od the altruists $\gamma$.

Consider ${\bf A}$:
\begin{equation}\label{e42}
{\bf A} = \left(
\begin{array}{cccccc}
-0.3 & 0 & 0 & 0.4 & 0 & 0\\
0 & -0.3 & 0 & 0 & 0.4 & 0\\
0 & 0 & -0.3 & 0 & 0 & 0.4\\
0 & 0 & 0.1 & 0 & 0.05 & 0\\
0.1 & 0 & 0 & 0 & 0 & 0.05\\
0 & 0.1 & 0 & 0.05 & 0 & 0\\
\end{array}\right),
\end{equation}
where $\alpha = -0.3,  \beta = 0.4, \sigma = 0.1,  \gamma = 0.05$.

\begin{figure}[h!]
	\begin{minipage}[h]{0.5\linewidth}
		\center{\includegraphics[width=1.1\linewidth]{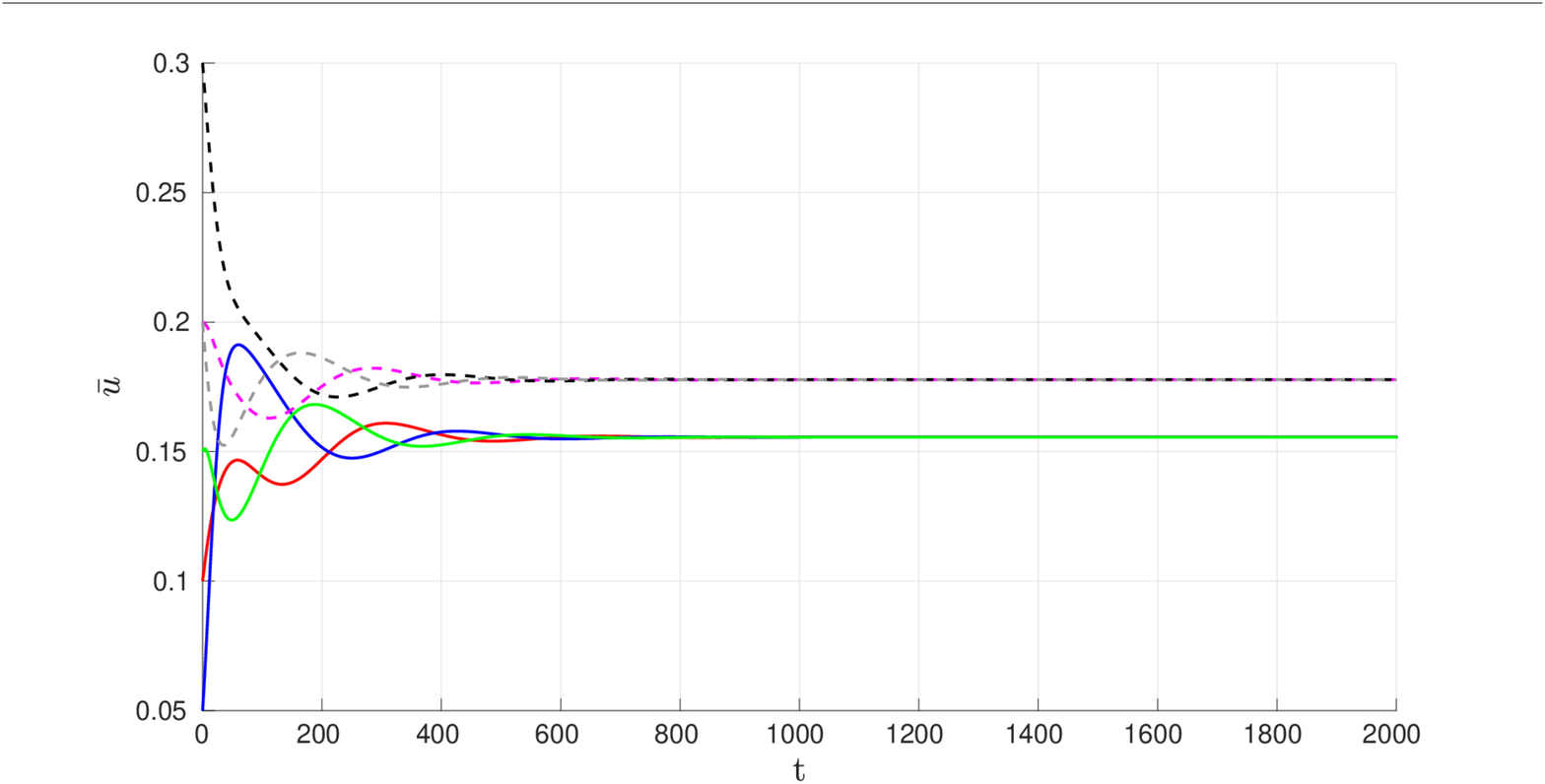}} a) \\
	\end{minipage}
	\hfill
	\begin{minipage}[h]{0.5\linewidth}
		\center{\includegraphics[width=1.1\linewidth]{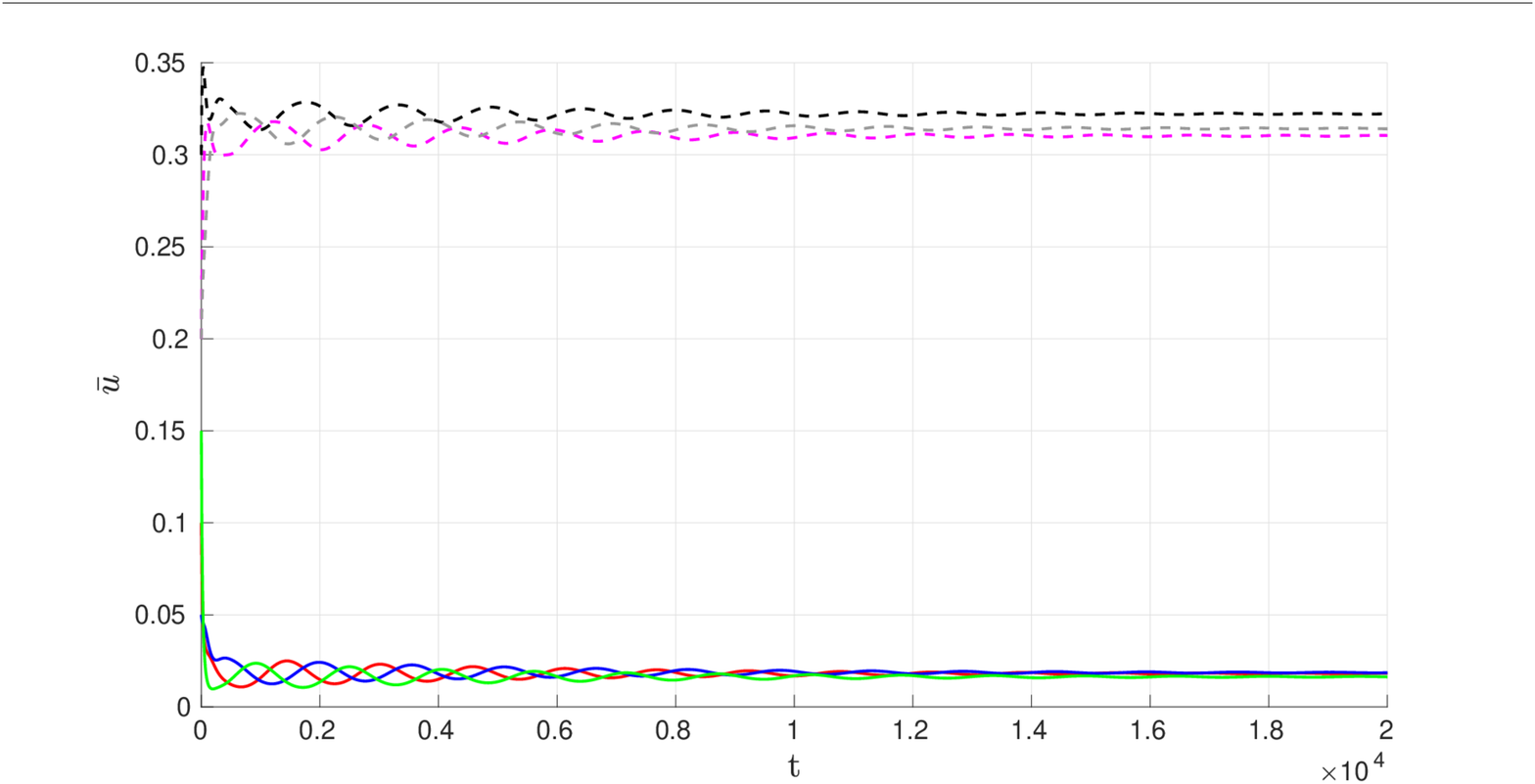}} \\b)
	\end{minipage}
	\vfill
	\begin{minipage}[h]{0.5\linewidth}
		\center{\includegraphics[width=1.1\linewidth]{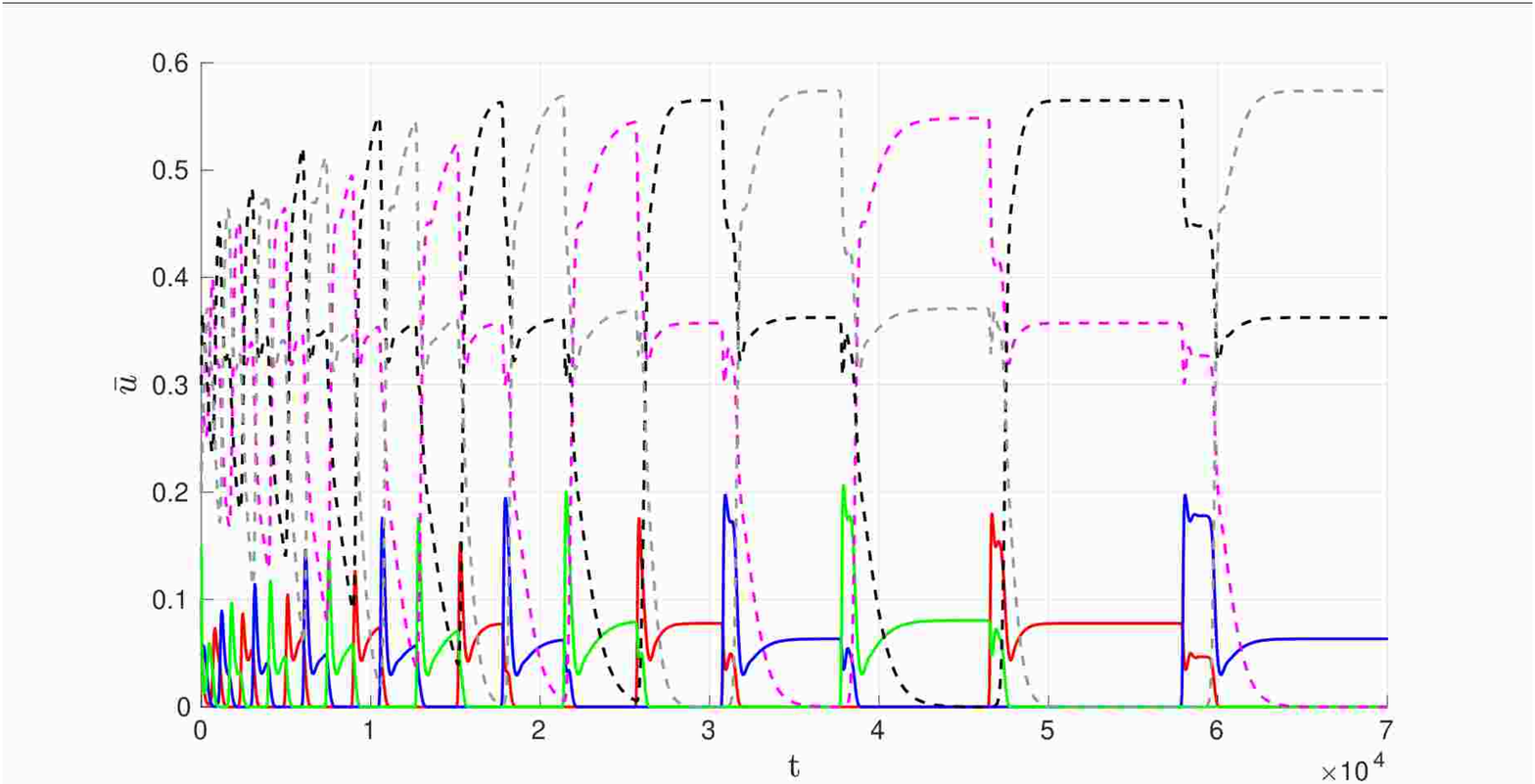}} c) \\
	\end{minipage}
	\hfill
	\begin{minipage}[h]{0.5\linewidth}
		\center{\includegraphics[width=1.1\linewidth]{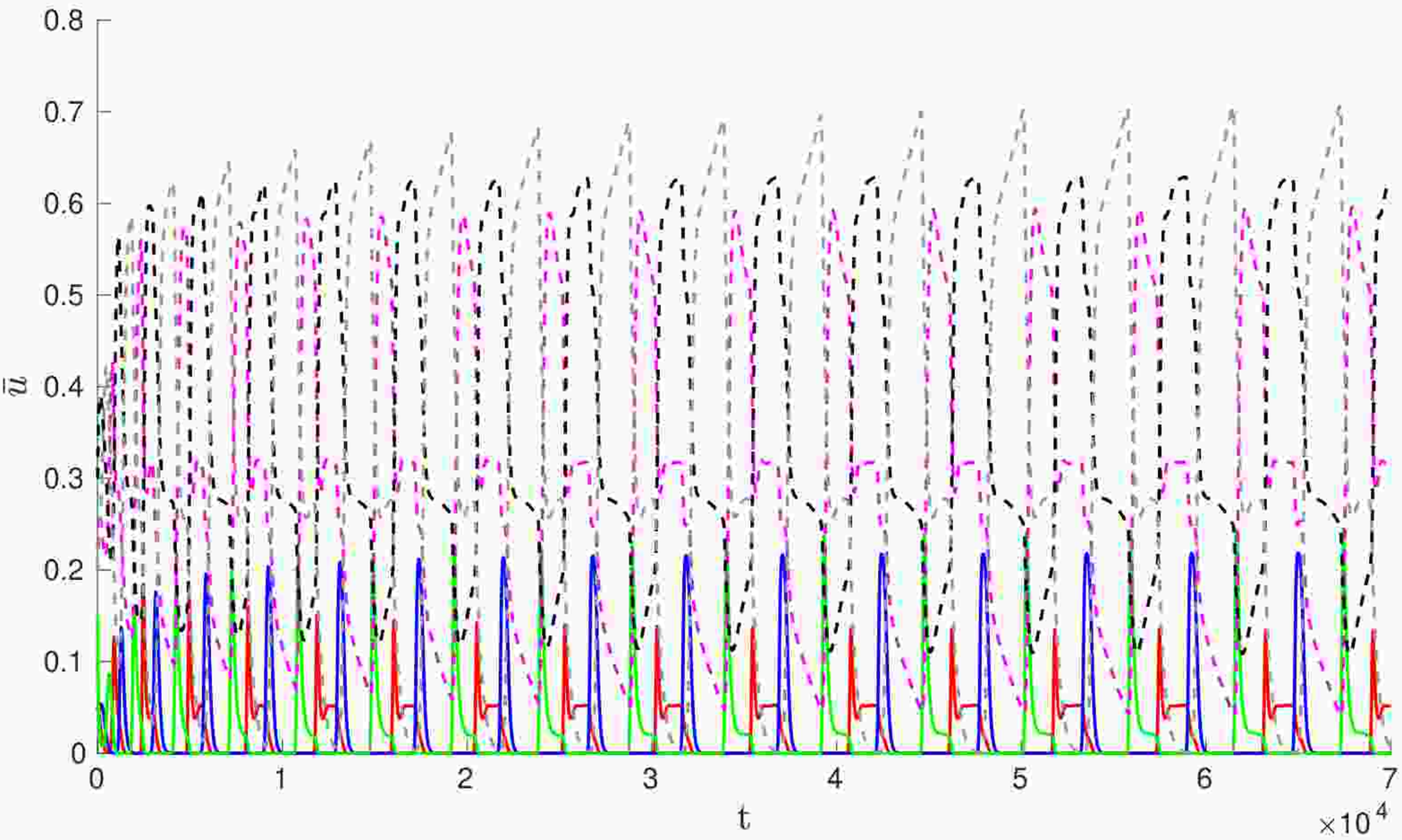}} d) \\
	\end{minipage}
	\caption{ The frequencies of the species in the hypercycle system (\ref{e41}) with matrix {\bf A} (\ref{e42}) changing over system time $t$:
		a) at the beginning, b) at the 125-th step of the fitness landscape evolutionary process, c) at the 175-th step of the fitness landscape evolutionary process, d) at the 200-th step of the fitness landscape evolutionary process.}
	\label{fig22}
\end{figure}

Fig. \ref{fig22} shows trajectories of altruists (dotted line) and egoists (solid line) at the beginning and 125, 175, 200 steps of the evolution \ref{e35} with ${\bf A}$ (\ref{e42}).

\section*{Conclusion}
In this paper, we applied an algorithm for the fitness landscape evolution of the replicator system.  We defined the limitation on the sum of squares of the fitness matrix coefficients while looking at the mean integral fitness maximum. We follow our previous study, suggesting that the evolutionary time of the hypercycle adaptation is much slower than the internal system dynamics time. 
The numerical simulation showed that the process of the mean fitness maximization is qualitatively similar for classical hypercycles and bi-hypercycles. At the beginning of the fitness landscape adaptation process, for a significantly long period in the evolutionary timescale,  the steady-state of the system (\ref{e1}, \ref{e2}) remains the same.  However, the structure of the transition matrix changes, which leads to the new transitions in the hypercycle system: besides the original connections, we get the backward cycle, autocatalysis, and new connections between species. This can be interpreted as a more diverse and sustainable evolutionary state. 
After some critical number of changes, the coordinates of the steady-state split into two parts: one species dominates, and its frequency converges to one, while the frequencies of the others converge to a minimum value. The latter process goes along with a significant increase in the autocatalytic coefficient for the dominant species, promoting its selfish behavior in the system.  According to the numerical investigation,  this dominant species is chosen by random and varies among the experiments. We suppose that this choice depends on computational errors.
As a final stage of the evolutionary process, there is a stabilization of the fitness landscape.  Here, calculations drastically depend on the restrictions on the steady-state coordinates. This process is similar to Eigen's error catastrophe proposed for the quasispecies systems.  The longevity of the evolutionary period before stabilization grows with the number of resources allowed in the system (\ref{e9}).

\newpage
\appendix

\section*{Appendix}\label{app}
Consider a sufficient condition for the extremum in the system observed. From (\ref{a1}), it follows: 
\begin{equation}\label{app1}
	(\delta^2 \bfA){\bf \bar{u}}+2(\delta \bfA)(\delta {\bf \bar{u}}) + \bfA (\delta^2{\bf \bar{u}}) = \delta^2\bar{f}\bf{I}.
\end{equation}
 Here, $\delta^2\bfA$ is the matrix with elements $a_{ij}''(\tau), i,j = 1,\ldots, n, $
$\delta^2{\bf \bar{u}}$ is the vector $(u_1''(\tau), \ldots, u_n''(\tau))$, $\delta^2f = {f}''(\tau).$

If $\delta \bar{f}(\tau) = 0,$ then from (\ref{a1}) we get 
$$
\bfA(\delta{\bf \bar{u}}) = -(\delta\bfA){\bf \bar{u}},
$$
or 
$$
(\delta {\bf \bar{u}}) = -\bfA^{-1}(\delta \bfA){\bf \bar{u}}.
$$
Let us substitute the latter into the expression (\ref{app1}), multiplying by ${\bf \bar{v}}$ from (\ref{a2}) and taking into account $(\delta^2{\bf \bar{u}}, \bf{I}) =0. $ Then, we have:
$$
\delta^2\bar{f} \left({\bf \bar{v}}, \bf{I}\right) = \left((\delta^2\bfA){\bf \bar{u}}, {\bf \bar{v}}\right) -2\left((\delta\bfA)\bfA^{-1}(\delta\bfA){\bf \bar{u}}, {\bf \bar{v}}\right).
$$
We denote $\bf{B}(\tau) = (\delta \bfA)\bfA^{-1}(\delta \bfA)$ --- the matrix with elements $b_{ij}(\tau).$ Then, the necessary condition for the extremum takes the form:
$$
\sum_{i,j=1}^{n}\left(a''_{ij}(\tau)-2b_{ij}(\tau)\right)u_iv_j<0, 
$$ 
if (\ref{e22}) reaches the equality.

\appendix

\section*{Acknowledgedments}
The work is supported by the Russian Science Foundation under grant 19-11-00008.
%The authors are grateful to  Artem S. Novozhilov for help, useful discussions
%and comments on the text.



\end{document}